\pdfoutput=1
\documentclass{aa}

\usepackage{txfonts} %For pretty font
\usepackage{amsmath, amssymb}
\usepackage[english=british]{csquotes}
\usepackage[T1]{fontenc}
\usepackage[utf8]{inputenc}
\usepackage{natbib}
\usepackage{lmodern}
\usepackage{graphicx,rotating}
\graphicspath{{.}{figures/}}
\usepackage{calc}
\bibpunct{(}{)}{;}{a}{}{,}
\usepackage{hyperref}
\usepackage[textwidth=3.5cm]{todonotes}
\usepackage{verbatim}
\usepackage{physics} %For equation type setting
\usepackage[]{siunitx} %For SI Units
\usepackage{multirow}
\usepackage{tabularx}
\usepackage{tikz-cd}

\usepackage{caption}
\usepackage{subcaption}

\usepackage{pstricks,pst-node}

\newcommand{\Omm}{\Omega_\mathrm{m}}

\newcommand{\LCDM}{$\Lambda$CDM}

\newcommand{\wcdm}{$w_0w_a$CDM}
\newcommand{\lcdm}{$\Lambda$CDM}

\newcommand{\gammat}{\gamma_\mathrm{t}}

\newcommand{\ds}{\Delta\Sigma}

\makeatletter
\renewcommand*\aa@pageof{page \thepage{} of \pageref*{LastPage}}
\makeatother

\hypersetup{pdfauthor={Sven Heydenreich}}

\begin{document}

\title{
Can dynamic dark energy explain the $S_8$ tension, the `lensing is low' effect, or strong baryon feedback?
}
\titlerunning{Dynamic dark energy and lensing}

\author{S.~Heydenreich$^{1}$, A.~Leauthaud$^{1}$, J.~DeRose$^{2}$
      }
\authorrunning{Heydenreich et al.}

\institute{
    $^{1}$ Department of Astronomy and Astrophysics, UCO/Lick Observatory, University of California, 1156 High Street, Santa Cruz, CA 95064, USA \\
    $^{2}$ Physics Department, Brookhaven National Laboratory, Upton, NY 11973, USA 
      }

\date{Version \today; received xxx, accepted yyy} % day month year

\abstract
{
% We investigate the impact of a dynamic dark energy cosmology on cosmological tensions related to weak gravitational lensing, namely the $S_8$ tension and the `lensing is low' effect. We compare lensing observables between a $\Lambda$CDM cosmology and a cosmology that employs the best-fit result of the Dark Energy Spectroscopic Instrument's DR2 analysis for the dark energy equation of state. We find that the galaxy-galaxy lensing signal is reduced by up to 7\% with respect to galaxy clustering; the suppression for cosmic shear reaches 14\%. While dynamic dark energy impacts the growth factor, both of these differences are primarily caused by changes to cosmological distance measures, which enter the lensing efficiency kernels. We further find that dynamic dark energy increases the thermal Sunyaev Zeldovich signal by about 15\%, but the signature of this increase does not explain the evidence for stronger baryon feedback that has been found in multiple recent studies.

We investigate the impact of a DESI motivated dynamic dark energy cosmology on three cosmological anomalies, the $S_8$ tension, the `lensing is low' effect, and observations of strong baryonic feedback. We analyze how these observations vary in $\Lambda$CDM versus dynamic dark energy. We find that the galaxy-galaxy lensing signal is reduced by up to 7\% with respect to galaxy clustering and that cosmic shear is suppressed by 14\%. These differences are primarily caused by changes to cosmological distance measures which enter the lensing efficiency kernels. In contrast, we find that dynamic dark energy increases the thermal Sunyaev Zeldovich signal by about 15\%, but that this is insufficient to substantially reduce the magnitude of baryonic effects. Thus, we find that dynamic dark energy may help explain two out of these three cosmological anomalies. DESI's dynamic dark energy has an important impact on cosmic expansion at $z\lesssim 0.5$, a regime where baryon acoustic oscillations are limited by the small volume. Because lensing is sensitive to distances, in addition to growth, we argue that lensing measurements are a promising alternative to constrain expansion deviations from $\Lambda$ at low redshifts.}


\maketitle

\section{Introduction}
\label{sec:introduction}

The {\LCDM} model is resoundingly successful in describing the structure and evolution of the Universe and has been the cosmological standard model for the last few decades. With the increasing accuracy and precision of cosmological observations, several tensions have emerged \citep[for a review, see][]{2022JHEAp..34...49A}. While actively debated, there appears to be growing evidence that the {\LCDM} framework may need to be revised.

In this work, we focus on one potential extension of the {\LCDM} model, the {\wcdm} model that heuristically allows the equation of state (EoS) of dark energy to vary as a linear function of the scale factor $w = w_0 + w_a (1-a)$.\footnote{We note that, while appearing to be a simple Taylor approximation to the `true' EoS, the choice of parametrization is physically motivated \citep{arXiv:gr-qc/0009008,2003PhRvL..90i1301L,arXiv:1505.05781}.} This is particularly motivated by recent analyses of the Dark Energy Spectroscopic Instrument (DESI), which found evidence for evolving dark energy in their flagship analyses of Baryon Acoustic Oscillations \citep{2025JCAP...02..021A,arxiv:2411.12022,arXiv:2503.14738}.\footnote{In particular, dark energy appears to have been phantom with $w<-1$ at earlier times and only recently crossed the phantom line \citep{arXiv:2503.14743}.} Interestingly, allowing for evolving dark energy appears to reduce tensions otherwise present in the constraint on the sum neutrino masses from DESI analyses \citep{arXiv:2503.14744}.

In this work, we investigate whether dynamic dark energy can also resolve other tensions present in analyses of the Universe's large-scale structure (LSS), in particular the lack of power present in measurements of the weak gravitational lensing effect. In recent years, analyses of cosmic shear surveys have precisely measured the matter power spectrum, yet these measurements consistently report a value of the matter clustering parameter, $S_8 = \sigma_8\sqrt{\Omega_\mathrm{m}/0.3}$, that is lower than the one inferred from the cosmic microwave background \citep{arxiv:2007.15632, 2023OJAp....6E..36D, Amon:2022, 2023PhRvD.108l3518L, arxiv:2304.00704, 2020A&A...641A...6P}. On scales smaller than those typically used for cosmological inference, weak gravitational lensing also appears to yield a lower signal amplitude when compared directly to galaxy clustering measurements, a phenomenon known as the `lensing is low' effect \citep{Leauthaud:2017,arxiv:2010.01143,2021MNRAS.502.2074L,2022MNRAS.509.1779L,2023MNRAS.520.5373L}.

It is important to note the possibility that these tensions are merely artifacts of systematic biases or incorrect analysis assumptions. The $S_8$-tension can be alleviated by allowing a more flexible prescription for the intrinsic alignment of galaxies \citep{arxiv:2407.04795}, and the most recent analysis by the Kilo-Degree Survey no longer sees a tension in the $S_8$ parameter \citep{2025arXiv250319441W}. Likewise, some studies argue that the `lensing is low' effect is caused by simplistic assumptions in the galaxy-halo connection model \citep{arxiv:2211.01744, 2023MNRAS.525.3149C}. In particular, there is a renewed interest in investigating the impact of baryon feedback on lensing observables, as direct observations indicate that this effect is stronger than previously assumed \citep{arxiv:2206.08591,arxiv:2403.20323,Bigwood:2024,Hadzhiyska:2024}.

In this work, we will investigate whether dynamic dark energy can explain these related tensions, namely the `lensing is low' effect and the $S_8$ tension. This is particularly motivated by \citet{2017MNRAS.471.1259J}, who found that cosmic shear prefers the same type of dynamic dark energy ($w_0>-1,\,w_a<0$) that the more recent DESI BAO results seem to favor. We will further research whether evidence for excess baryon feedback can also be releated to a preference for dynamic dark energy.

Throughout this work, we will assume the best-fit parameters for the dark energy EoS from the DESI-DR2+CMB+Union3 BAO analysis, namely $w_0=-0.67$, $w_a=-1.09$, $H_0=65.9$, and $\Omega_\mathrm{m}=0.33$, and investigate how cosmological quantities relate to a {\LCDM} cosmology with the same cosmological parameters (apart from setting $w_0=-1,\,w_a=0$). We will call the former cosmology the `DESI-BAO' cosmology, the latter will be denoted by the {\LCDM} cosmology.

% \begin{itemize}
% \item For decades {\lcdm} has been the preferred model, with no conclusive indication of lambda budging from $-1$. As such, many anlaysis by default assume $w=-1$
% \item However, DESI D1 and DR2 show increasing evidence for dynamical dark energy. Although DESI alone is consistent with LCDM, DESI in combination with other data sets has this preference for wo and wa.
% \item Once the idea is dynamicla dark energy is cracked open, it is interesting to think about what other mysteries this might solve, and what other smoking guns might already exits in the data. Combined evidence. For example, as hilight in the DESI neutrino paper, DESI cosntarins on neutrinos in LCDM have posteriors that peak at negative masses which is unphysical and inconsistenw tih neutrino experiments. However, when analyze in wcdm, things become more conssitent.
% \item In this paper, we investigate to what extent wCDM might also impact two other cosmological puzzles. THe first is lensing is low. Define what this is. Alexie can write
% \item The second are constraints on baryonic effects. Describe here. Also state wether or not 
% \item The goal of this paper is not to carry out a full analysis of these two effects, but rather to investigate if dynaimcal DE has the potential to "solve" these two cases in one fell swoop.
% \end{itemize}
%     DESI's recent BAO measurements favor a $w_0w_a$CDM cosmology at 2.5-3.5$\sigma$. We investigate if a $w_0w_a$CDM model would resolve other known tensions, such as the evidence for strong baryon feedback and the lensing is low effect.

\section{Theoretical background}
\label{sec:theory}

\subsection{Cosmology}
We assume a Friedmann-Lemaître-Robertson-Walker metric with curvature
$K=0$ and describe dark energy with the Chevallier-Polarski-Linder
parameterization $w(a)=w_0+w_a(1-a)$
\citep{2003PhRvL..90i1301L}. Comoving radial distances follow
\begin{equation}
\chi(z)=c\int_{0}^{z}\frac{\dd z'}{H(z')}\; ,
\end{equation}
where
$H^2(z)=H_0^2\bigl[\Omm(1+z)^3+\Omega_{\mathrm{DE}}(z)\bigr]$ and
$\Omega_{\mathrm{DE}}(z)=\Omega_{\mathrm{DE},0}(1+z)^{3(1+w_0+w_a)}
\exp\!\bigl[-3w_a z/(1+z)\bigr]$.
Angular diameter distances follow
$D_A=\chi/(1+z)$ and enter all projected observables. Linear growth
factors $D_+(z)$ are obtained by solving the growth equation with the
above $H(z)$ \citep{1999astro.ph..5116H}.
Throughout we use the {\tt Eisenstein \& Hu} transfer function
\citep{1998ApJ...496..605E} and the
{\tt HMcode2020} prescriptions for non-linear matter power spectra.

\subsection{Galaxy--halo connection}
We model the relationship between galaxies and matter with the halo
model \citep{2002PhR...372....1C}. For central and satellite galaxy
occupancy we adopt a five-parameter HOD of the form introduced by
\citet{2005ApJ...633..791Z}, truncated at the virial radius and
assuming an NFW density profile for satellites.

\subsection{Projected two-point statistics}
We model the projected two-point statistics for cosmic shear ($\kappa\kappa$), galaxy-galaxy lensing ($\delta_g\kappa$) and projected clustering ($\delta_g\delta_g$) via their projected power spectra $C_\ell$, which can be derived from the 3-dimensional matter power spectrum $P$ via the following relations \citep[e.g$.$][]{Krause:2017, Limber:1953, arxiv:2012.08568}:
\begin{align}
\label{eq:C_ells}
    C^{ij}_{\kappa \kappa}(\ell) ={}&{} \int d\chi \frac{q^i_\kappa(\chi) q^j_\kappa(\chi)}{\chi^2} P\left(\frac{\ell+\frac{1}{2}}{\chi}, z(\chi)\right),
\nonumber\\
    C^{ij}_{\delta_g \kappa}(\ell) ={}&{} \int d\chi \frac{q^i_\delta\left(\frac{\ell+\frac{1}{2}}{\chi},\chi \right) q^j_\kappa(\chi)}{\chi^2} P\left(\frac{\ell+\frac{1}{2}}{\chi}, z(\chi)\right), 
\nonumber\\
    C^{ij}_{\delta_g \delta_g}(\ell) ={}&{} \int d\chi \frac{q^i_\delta\left(\frac{\ell+\frac{1}{2}}{\chi},\chi \right) q^j_\delta\left(\frac{\ell+\frac{1}{2}}{\chi},\chi \right)}{\chi^2} P\left(\frac{\ell+\frac{1}{2}}{\chi}, z(\chi)\right)\ ,
\end{align}
where $i$ and $j$ denote different combinations for tomographic redshift bins, which we will omit going forward. In the equations above, $\chi$ denotes the comoving radial distance, $q_\kappa$ is the lensing efficiency kernel, and $q_\delta$ the radial weight function. They are given by
\begin{eqnarray}
 q_\kappa (\chi) &=& \frac{3H_0^2 \Omega_m\chi}{2a(\chi)} \int_{\chi}^{\chi_h} d\chi^\prime \left(\frac{\chi^\prime - \chi}{\chi}\right) n_\kappa(z(\chi^\prime)) \frac{\dd z}{\dd\chi^\prime} , \nonumber \\
 q_\delta (\chi) &=& b\,  n_\delta(z(\chi)) \frac{\dd z}{\dd\chi}\ ,
 \label{eq:lensing_efficiencies}
\end{eqnarray}
where $n(z)$ denotes the redshift distribution of lenses (for $q_\delta$) and sources (for $q_\kappa$), and $b$ is the linear galaxy bias.\footnote{Technically, the galaxy bias is a function of $k$ and $z$. Here we only compare theoretical predictions between cosmologies; hence, we set $b$ to be constant.} Configuration-space correlation functions
$\xi_\pm(\vartheta),\,\gamma_t,\,w_\theta$ are obtained through the methods outlined in \citet{Stebbins:1996} and optimized for numerical integration in \citet{arxiv:2012.08568}. In our simulation-based analysis, we also utilize the excess surface density $\Delta\Sigma$, which is related to $\gammat$ via \citep{2001PhR...340..291B}
\begin{eqnarray}
    \gammat(\theta) &=& \Delta\Sigma\left(\theta\frac{\chi_\mathrm{l}}{(1+z_\mathrm{l})}\right) \, \Sigma_\mathrm{crit}^{-1}(\chi_\mathrm{l},\chi_\mathrm{s}), \notag\\
    \Sigma_\mathrm{crit}^{-1}(\chi_\mathrm{l},\chi_\mathrm{s}) &=& \frac{4\pi G}{c^2}\frac{(1+z_\mathrm{l})(\chi_\mathrm{s}-\chi_\mathrm{l})}{\chi_\mathrm{s}}\mathcal{H}(\chi_\mathrm{s}-\chi_\mathrm{l}) ,
\end{eqnarray}
where $\chi_\mathrm{l,s}$ are the comoving radial distances to the lens and source, respectively, and $\mathcal{H}$ is the Heaviside step function.

\subsection{Thermal SZ signal and the CAP filter}
Thermal Sunyaev-Zeldovich (tSZ) temperature fluctuations are sourced
by the Compton-$y$ field,
$y\propto\int\!P_e\,\dd l$ \citep{1972CoASP...4..173S}.  We follow
\citet{2015ApJ...808..151G} and define the compensated aperture
photometry (CAP) filter
\begin{align}
    y_{\mathrm{CAP}}(\theta)= {}&{}\int\!\dd^2\vartheta\,
    y(\vartheta)\,U_{\theta_\mathrm{ap}}(|\boldsymbol{\vartheta}|)
    ,\notag\\
    U_{\theta_\mathrm{ap}}(\vartheta)= {}&{} \quad 
    \frac{1}{\pi\theta_\mathrm{ap}^2}\!
    \left[\,1-\Theta(\vartheta-\theta_\mathrm{ap})\right] \notag\\
    {}&{} -\frac{1}{\pi(\sqrt{2}\,\theta_\mathrm{ap})^2}\!
    \left[\,1-\Theta(\vartheta-\sqrt{2}\,\theta_\mathrm{ap})\right],
\end{align}
which removes the large-scale $y$-mean and facilitates comparison with
observations and simulations.

\subsection{Used data products}
We employ the publicly released lens and source redshift distributions
from
DESI-DR1 Bright Galaxy Survey \citep{2023AJ....165..253H},
DESI-DR1 Luminous Red Galaxies \citep{2023AJ....165...58Z},
KiDS-1000 \citep{2021A&A...647A.124H},
DES Year-3 \citep{2021MNRAS.505.4249M},
and HSC-SSP Year-3 \citep{2023MNRAS.524.5109R}.
All $n(z)$ histograms are re-binned to $\Delta z=0.005$ and normalised
to unity before entering any calculation.

\subsection{Used simulations}
    Non-linear predictions are validated against the
    {\sc AbacusSummit} \citep{2021MNRAS.508.4017M}
    suite, generated with the high-precision {\sc Abacus} $N$-body code
    \citep{2021MNRAS.508..575G}.  We analyse boxes
    \texttt{base\_c000\_ph000} (\lcdm) and
    \texttt{base\_c111\_ph000} (\wcdm) using
    the public {\tt TabCorr} package \citep{Neistein:2011,Zheng:2016,Lange:2019a,Lange:2019b} to tabulate $w_p$ and $\Delta\Sigma$.

\section{Methods}
\label{sec:methods}
    We perform two kinds of analyses. Our first approach is a theoretical investigation utilizing the public python library \texttt{pyccl}\footnote{https://github.com/LSSTDESC/CCL} \citep{arXiv:1812.05995}. Since we are investigating percent-level effects and power spectrum predictions in dynamic dark energy scenarios are not extensively validated, we supplement this investigation with a simulation-based analysis utilizing \texttt{AbacusSummit} and \texttt{TabCorr}. While the latter does not provide a perfect match to the DESI best-fit cosmology, it provides a stencil derivative cosmology with the same parameters as the base cosmology, except setting $w_0=-0.9$ and $w_a=-0.4$. While not close to the best-fit DESI cosmology, it deviates from {\lcdm} in the same direction as the DESI best-fit $w_0w_a$CDM results. We can thus test whether we qualitatively recover the same trends as in the theoretical investigation when comparing this simulation to the base, $\Lambda$CDM one.

    \subsection{Theoretical investigation}
        We assume the best-fit values of the DESI $w_0w_a$CDM cosmology ($w_0=-0.64$, $w_a=-1.27$) and compare different cosmological quantities and observables between this and a fiducial $\Lambda$CDM cosmology with the same cosmological parameters.

        As a first step, we investigate how dynamic dark energy impacts the individual components that source the final cosmological observables. All LSS observables are impacted by the growth factor, which describes the linear growth of structure and parametrizes the scale-independent evolution of the matter power spectrum over cosmic time. Further, distance measures are also impacted by the Universe's expansion history. They enter all LSS calculations when one converts between angular and phyiscial distances of configuration-space statistics. Crucially, for weak lensing, they additionally enter the calcultions in prefactors quantifying the lensing efficiency $q_\kappa$.
        
        Our next step is to compute the projected angular power spectra $C_\ell$ for quantities of interest, here galaxy clustering, galaxy-galaxy lensing, cosmic shear, and tSZ. We assume a linear bias of $b=1$ for all tracers for simplicity. We subsequently predict the projected galaxy clustering correlation function $w_\theta$, the tangential shear around galaxies $\gamma_\mathrm{t}$, and the cosmic shear 2-point correlation functions $\xi_\pm$. We use public redshift distributions of DES-Y3, KiDS, and HSC-Y3 for the source galaxy populations. We further transform the projected tSZ power spectrum into predictions for the compensated aperture (CAP) filter, as used in \citet{Hadzhiyska:2024} and other works. For the lens galaxies, we utilize the redshift distributions of DESI-DR1 BGS and LRG. 
    
        We then use a matched filter approach, as detailed in \citet{Leauthaud:2022} to project the predictions from the DESI-BAO cosmology into a single amplitude, where we take the $\Lambda$CDM prediction as the reference data vector and use a covariance for the respective lens-source bin combination as detailed in \citet{Yuan:2024}. For all purposes, we assume number densities and redshift distributions for the sample of lens galaxies described in \citet{Heydenreich:2025}.
        We then compare the amplitude of the different signals in the DESI-BAO cosmology to compare how a deviation from $\Lambda$ impacts different LSS tracers respectively.

    \subsection{Simulation-based analysis}
        The theoretical investigation detailed above faces the caveat that codes modeling nonlinear structure formation have not been extensively validated for non-$\Lambda$CDM cosmologies. We thus supplement our analysis by an investigation of the \texttt{AbacusSummit} simulation suite, which provides a simulation that deviates from the base cosmology with $w_0=-0.9$ and $w_a=-0.4$. While far away from the DESI best-fit cosmology, the deviation from {\lcdm} is in a similar direction as the DESI best-fit {\wcdm} results, so we can test whether we qualitatively recover the same trends as in the theoretical investigation when comparing this simulation to the base, $\Lambda$CDM one.

        We tabulate the projected correlation function $w_p$ and the excess surface density $\Delta\Sigma$, which are related to the $w_\theta$ and $\gamma_\mathrm{t}$ statistics above, using the public \texttt{TabCorr} code. We then fit an HOD \citep{2005ApJ...633..791Z} to an observed $w_p$ clustering signal from \citet{Leauthaud:2017} using the \texttt{nautilus} sampler \citep{Lange:2024}, and optimize the fit by performing a scalar minimization of the likelihood at the best-fit cosmology determined by \texttt{nautilus}. Using these best-fit HOD parameters, both for the {\lcdm} and DESI-BAO cosmology, we then predict the projected clustering $w_p$ (to check consistency) and the excess surface density $\Delta\Sigma$. To compare absolute values, we again use the matched filter approach to compress the DESI-BAO predictions into a single amplitude, given the {\lcdm} predictions as the reference data vector.

\section{Results}
\label{sec:results}
    \subsection{Cosmological quantities}
\begin{figure}
            \centering
            \includegraphics[width=\linewidth]{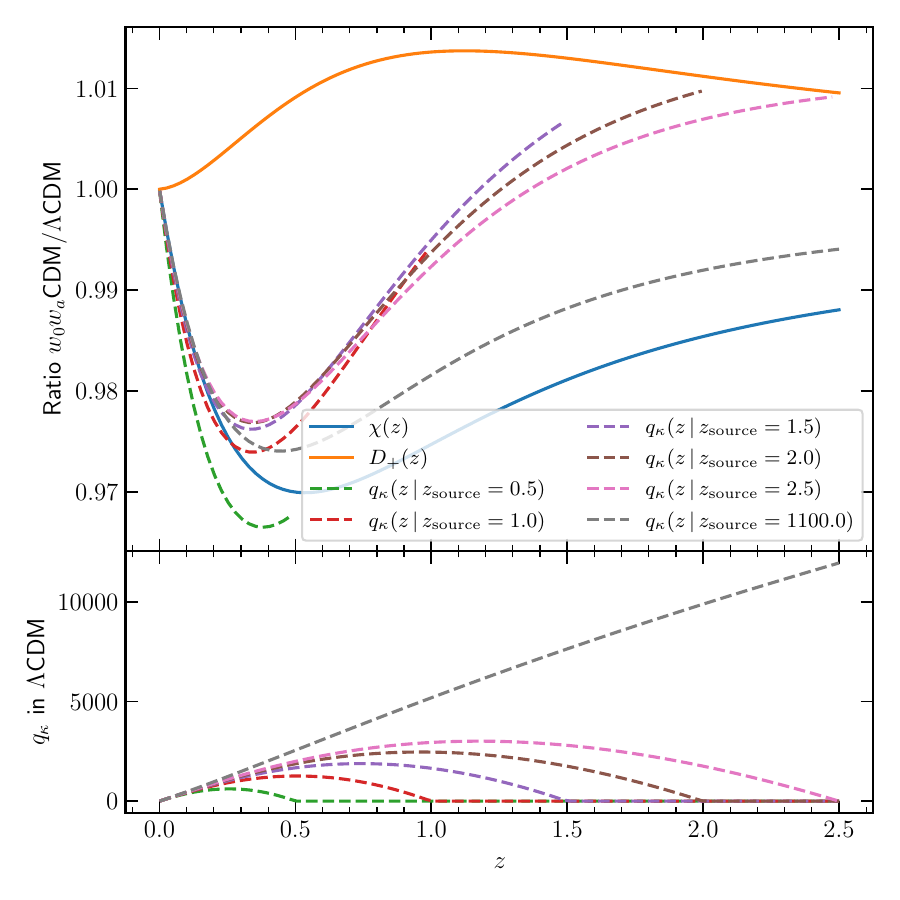}
            \caption{Top: Ratio of cosmological distances (blue), the normalized growth factor (orange), and lensing efficiencies (dashed lines) under a DESI-BAO model, normalized by the values in the corresponding {\lcdm} model. Bottom: The lensing efficiencies $q_\kappa$ for different source redshifts in {\lcdm}.}
            \label{fig:distance_ratios}
        \end{figure}

        Dynamic dark energy changes the expansion history of the Universe, and thus impacts both growth factors and distance measures. It is therefore plausible that the conversion from angular to physical coordinates, under the assumption of $\Lambda$CDM, could induce biases. As can be seen in Fig.~\ref{fig:distance_ratios}, the effect caps at about 3\% for $z\sim 0.8$. While certainly relevant for precision cosmology (e.g.~when measuring $a_\perp$ and $a_{||}$ in BAOs), physical distances differing by 3\% alone is unlikely to explain the effects seen in \citet{Leauthaud:2017,2021MNRAS.502.2074L,2022MNRAS.509.1779L} or in \citet{Hadzhiyska:2024}. We can see in the same figure that the growth factor is about 1.5\% larger at $z\sim 1$ in the DESI-BAO cosmology compared to {\lcdm}. Crucially, we find that the lensing efficiency $q_\kappa$ drops by about 2.5\% at $z\sim 0.4$. As this factor does not contribute to clustering observables, but contributes to galaxy-galaxy lensing observables and, squared, to cosmic shear observables, we expect a relative suppression of these statistics with respect to clustering observations.
    \subsection{Impacts on observables: Theoretical investigation}
        Dynamic dark energy impacts structure formation in a non-trivial way. We assess the ratio of $C_\ell$ for projected quantities of interest, here galaxy clustering, galaxy-galaxy lensing, cosmic shear, and tSZ.

\begin{figure*}
            \centering
            \includegraphics[width=\linewidth]{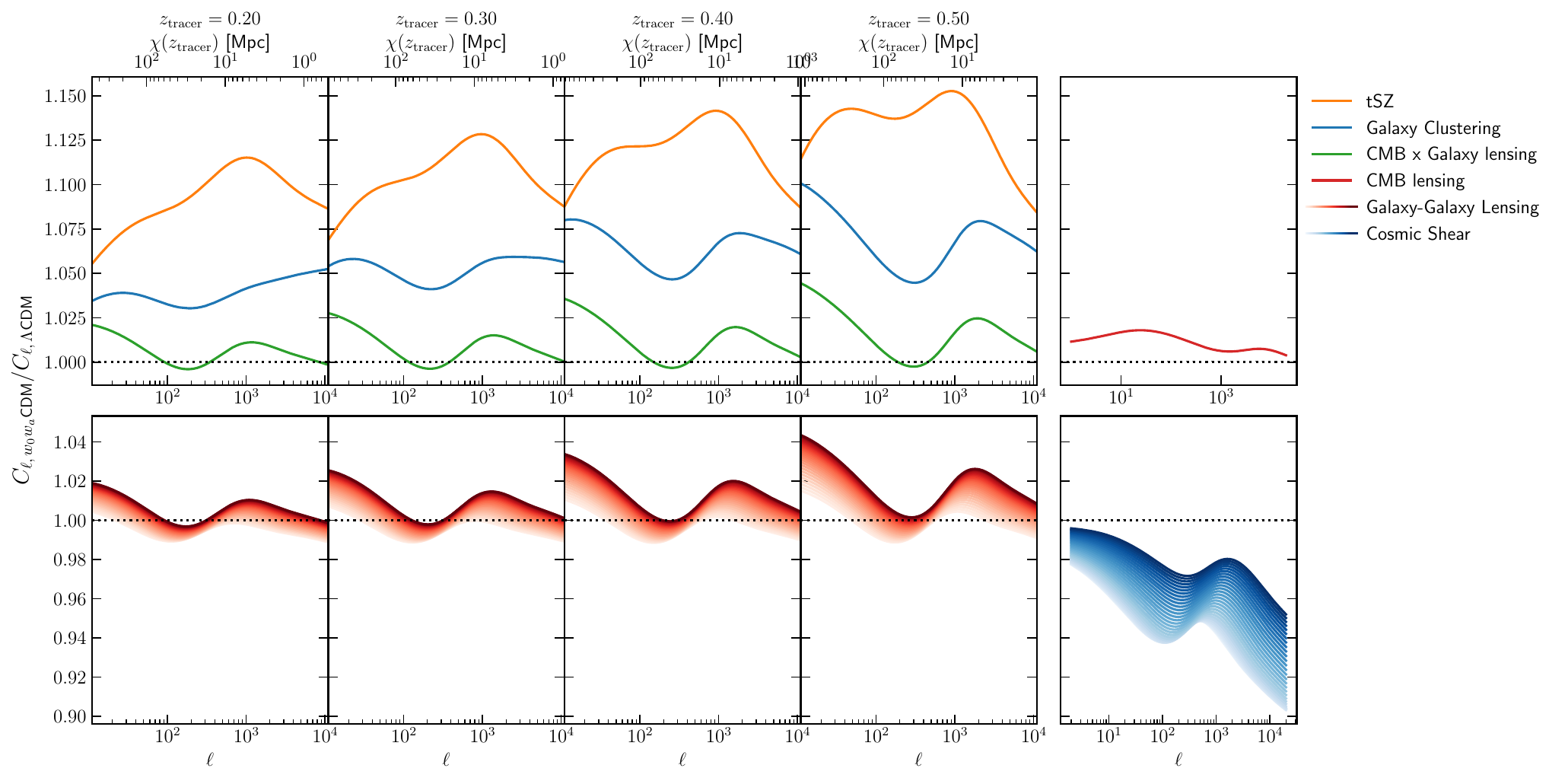}
            \caption{Ratio of projected angular power spectra $C_\ell$ under a DESI-BAO cosmology vs a {\lcdm} cosmology. The left four columns represent different redshifts of tracer galaxy populations. The right-most column shows quantities that do not depend on tracer populations (CMB lensing and cosmic shear). The top row shows the tSZ, galaxy clustering, and CMB x Galaxy lensing signal ratios for the tracer galaxy populations in the left columns, and the CMB lensing signal ratio in the right column. The bottom row on the left shows the galaxy-galaxy lensing signal ratio for different source redshifts, where the tracer galaxy population constitutes the lens galaxies. On the right, we show the cosmic shear signal ratio. The color coding for the galaxy-galaxy lensing and cosmic shear signals indicates source redshift distributions varying between $z=0.3$ (light colors) and $z=1.5$ (dark colors).}
            \label{fig:cell_ratios}
        \end{figure*}

        As a first investigation, we estimate the impact of a DESI-BAO cosmology by predicting projected angular power spectra for galaxy clustering, galaxy-galaxy lensing, cosmic shear, tSZ, and CMB lensing. To simplify our assumptions, we model the redshift distributions as Gaussians of width $\sigma=0.2$; the lens redshifts are centered around $z\in \{0.2,0.3,0.4,0.5\}$, the source redshifts vary between $0.3 \leq z \leq 1.5$. We then compare the ratio of the $C_\ell$ for the DESI-BAO cosmology to the {\lcdm} cosmology in Fig.~\ref{fig:cell_ratios}. We can observe a few general trends:
        \begin{itemize}
            \item Projected galaxy clustering is increased in the DESI-BAO cosmology. The amount of increase increases with redshift from 5\% at $z=0.2$ to 10\% at $z=0.5$. This is caused both by the increase in the growth factor $D_+$, and by a decrease in the comoving distance $\chi$, which increases $\dd z/\dd\chi$ in $q_\delta$ (Eq.~\ref{eq:lensing_efficiencies}).
            \item Galaxy-galaxy lensing increases with both source- and lens-redshift in the DESI-BAO cosmology with respect to {\lcdm}. For low source- and lens-redshift we observe a slight decrease of the GGL signal, whereas for high lens- and source-redshifts the ratio is slightly above unity. Here, the increase in the growth factor gets countered by a general decrease of $q_\kappa$. This is most effective when the lens galaxy populations are in the region where $q_\kappa$ is suppressed most ($z\sim 0.35)$. With increasing source redshift, the suppression of $q_\kappa$ decreases, and the region where $q_\kappa$ is suppressed the most no longer strongly contributes to the lensing signal, as the absolute value of $q_\kappa$ around $z\sim 0.35$ becomes very small.
            \item Cosmic shear is decreased in the DESI-BAO cosmology by about 12\% for low source redshifts and a few per-cent for high source redshifts. The decrease of cosmic shear is stronger than for GGL, as $q_\kappa$ enters the equation twice. The amount of signal suppression depends both on the amount that $q_\kappa$ drops, as well as on the value of $q_\kappa$ around that drop. For high source redshifts, the region around $z\sim 0.35$ does not contribute substantially to the lensing signal, meaning that the lower lensing efficiency at these redshifts is no longer as relevant.
            \item The CMB lensing signal is relatively unaffected by the change in cosmology. While the $q_\kappa$ drops strongly for the CMB lensing signal, the CMB lensing efficiency is practically negligible at $z<0.5$, meaning that this drop does not substantially impact the lensing signal, and gets counter-acted by the increased growth rate.
            \item The excess in tSZ signal increases with lens redshift, from about 10\% at $z=0.2$ to 15\% at $z=0.5$. This is likely sourced by the non-linear relationship between halo mass (which in turn relates to the growth factor) and tSZ signal, with tSZ $\propto M^\frac{5}{3}$.
        \end{itemize}
        
        Overall, we see that summary statistics in the DESI-BAO cosmology behave in a way that is consistent with the observed `lensing is low' trend -- the GGL signal is suppressed by 5\%-15\% with respect to the galaxy clustering, depending on source and lens redshift, and the cosmic shear signal is suppressed by up to 30\% with respect to galaxy clustering. This could explain both a lower $S_8$ value measured by cosmic shear surveys and the more direct lensing is low effect, which describes a lack of power of the GGL signal with respect to galaxy clustering.

        Regarding excess baryon feedback, the conclusion is a bit less unanimous. While we do predict an increased tSZ signal around galaxies by about 15\% in the DESI-BAO cosmology, this is not necessarily equivalent to an extended baryon feedback scenario. As can be seen in \citet{Hadzhiyska:2024}, at least in the kSZ, the extended baryon feedback signal manifests as a decrease of the signal at smaller scales, where the baryons are expelled from the halo. 

        \begin{figure}
            \centering
            \includegraphics[width=\linewidth]{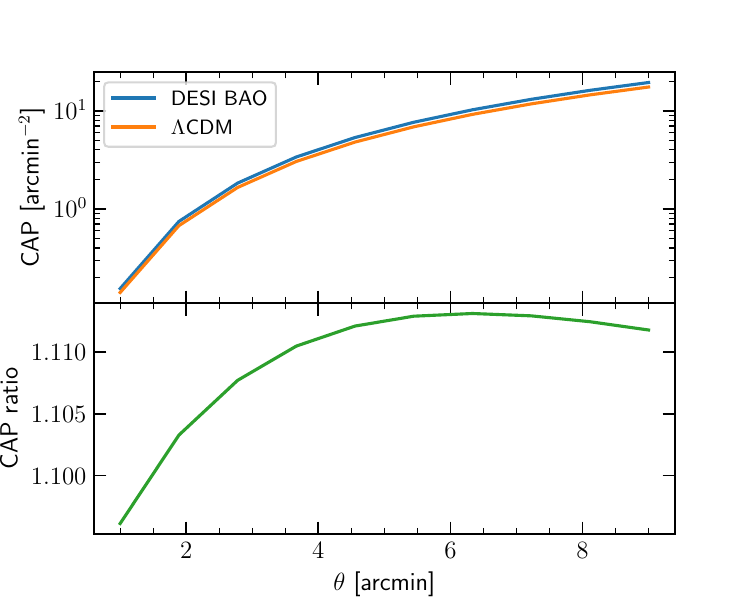}
            \caption{Projected tSZ signal around galaxies in a DESI-BAO cosmology compared to {\lcdm} using the more commonly used compensated aperture filter (CAP) technique. While the total tSZ signal increase is 15\%, strong baryon feedback is characterized by a signal drop on small scales relative to the large scales. This drop is present, but observed baryonic effects are much larger.}
            \label{fig:cap_ratios}
        \end{figure}

        We note that DESI shows a large degeneracy along the $w_0$-$w_a$ axis, so that stronger deviations from {\lcdm}, and thus also a stronger suppression of GGL and cosmic shear with respect to galaxy clustering, are definitely plausible.

    \subsection{A realistic case: Configuration space statistics with DESI x HSC, KiDS, DES and ACT}
\begin{figure*}
            \centering
            \includegraphics[width=0.32\linewidth]{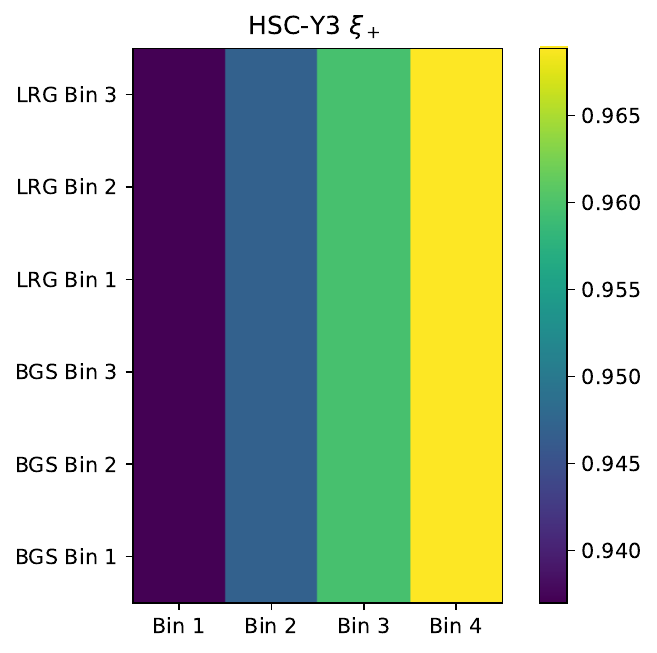}
            \includegraphics[width=0.32\linewidth]{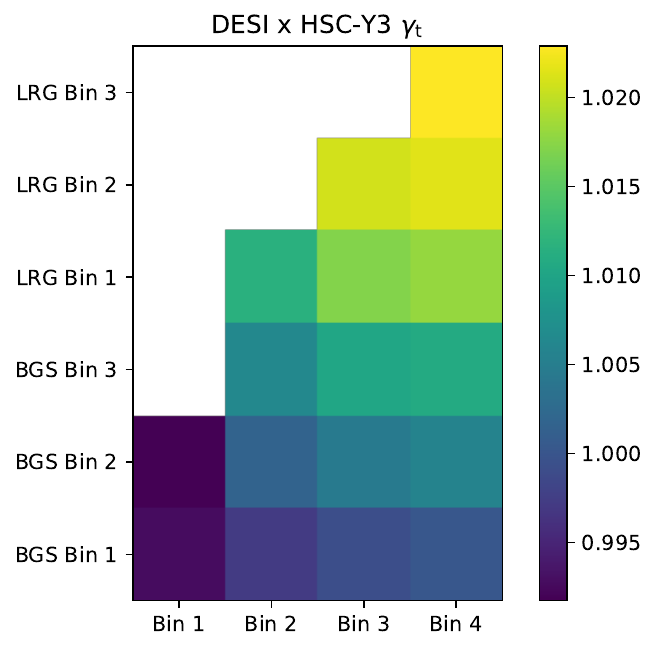}
            \includegraphics[width=0.32\linewidth]{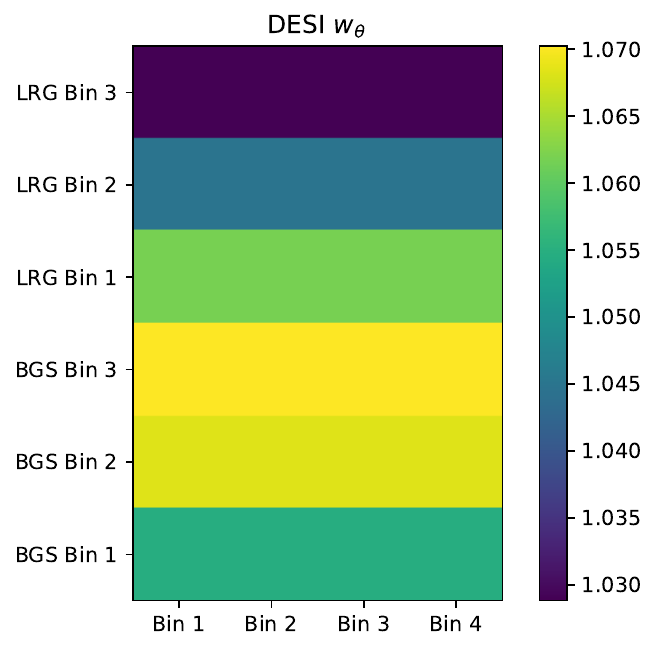}
            \\
            \includegraphics[width=0.32\linewidth]{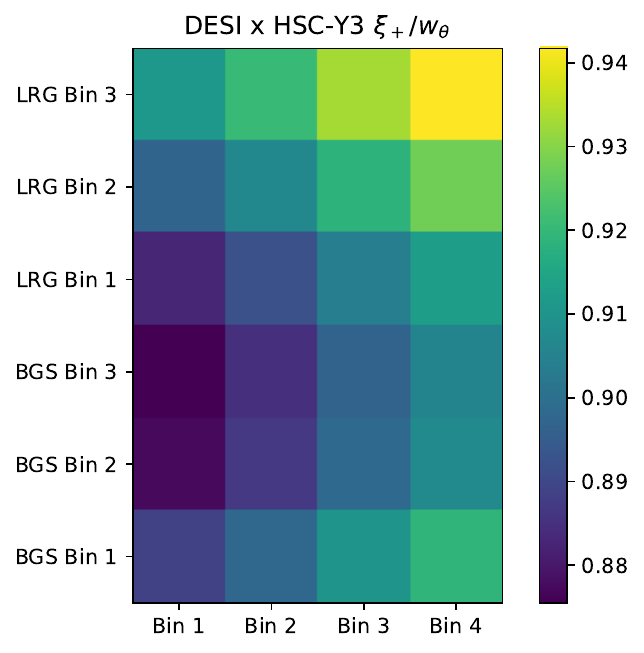}
            \includegraphics[width=0.32\linewidth]{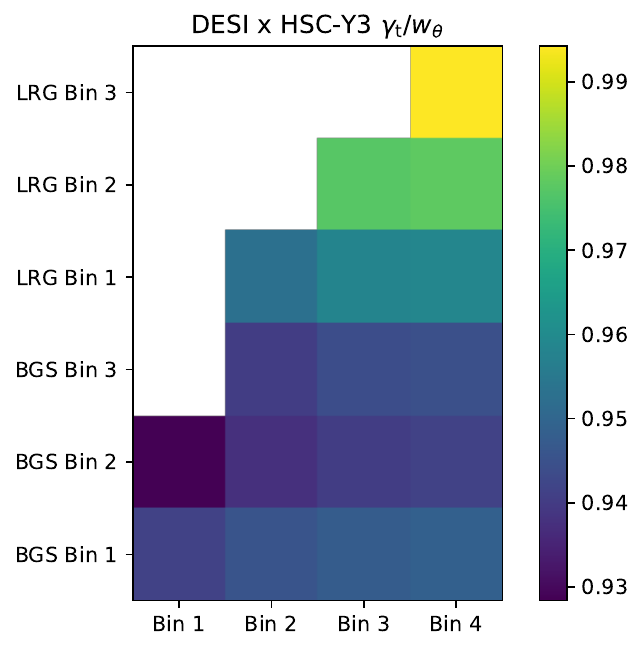}
            \caption{Top row: Predicted amplitudes of projected clustering, tangential shear, and cosmic shear in the DESI-BAO cosmology compared to {\lcdm} for DESI-DR1 lenses and HSC-Y3 source galaxies. Bottom row: Ratios of the lensing to clustering amplitudes in the DESI-BAO cosmology divided by the same ratios in {\lcdm}.}
            \label{fig:realistic_amplitudes}
        \end{figure*}

        Instead of general redshift trends, we now investigate a realistic scenario of GGL and projected clustering measurements using DESI-DR1 lens galaxies and source galaxies from the three major Stage-III lensing surveys. As above, we compare the predictions of CCL in the DESI-BAO cosmology and {\lcdm}. We utilize the redshift distributions of the measured tracers and transform the predicted power spectra to their configuration space equivalents. To better compare measurements, we project the DESI-BAO predictions to a single amplitude using the matched filter approach, where we take covariance estimates computed as in \citet{Yuan:2024}, and the reference data vector is the {\lcdm} prediction. In Fig.~\ref{fig:realistic_amplitudes} we show the results for cross-correlations between DESI and HSC-Y3. A clear picture emerges, which confirms the theoretical investigations above: The GGL signal is suppressed, with respect to the projected clustering signal, by up to 7\%, where the suppression is largest for small lens-source separations. The cosmic shear signal is suppressed by up to 13\% with respect to the projected clustering signal, where the suppression is strongest for high lens and small source redshifts. Interestingly, the fact that this suppression increases when the lens-source distance decreases, implies that an effect like this can potentially be masked by an uninformative intrinsic alignment model.

    \subsection{Simulation-based analysis}
        To validate our theoretical investigations, we perform a simulation-based analysis by comparing $\ds$ signals between a {\LCDM} and a DESI-BAO-like cosmology, where the HODs that model the $\ds$ signals are fitted to the observed projected clustering signal.

        We find that the predicted projected clustering signals are consistent between {\lcdm} and the DESI-BAO cosmology, with the DESI-BAO prediction showing an amplitude of 0.998. As the HODs were explicitly tuned to achieve consistent clustering, this is not surprising at all, but serves as an important validation check to confirm that fitting accuracy and flexibility do not impact our results. In contrast, the excess surface density shows a 2\% decrease, exhibiting an amplitude of 0.978. While not a strong effect, linearly extrapolating this to the best-fit cosmology of the DESI-DR2 analysis, it matches with the predictions from the theoretical investigations above, making a strong case for their validity.

    \section{Discussion}
    \label{sec:discussion}
    \subsection{Dynamic dark energy and the lensing is low effect}
        The lensing is low effect describes a discrepancy between the amplitude of the projected clustering signal and the GGL signal. This discrepancy is most pronounced in the KiDS survey, where we find that the DESI BAO cosmology lowers the GGL signal by 15\% compared to the projected clustering signal. A dynamic dark energy scenario at the best-fit cosmology of DESI-DR1 would account for about half of that discrepancy, as we observe an up to 10\% suppression of the GGL signal. As an additional benefit, the $S_8$ tension would also be alleviated, as the cosmic shear signal is suppressed by 10\% to 15\% with respect to the projected clustering signal. While the DESI-DR1 best-fit result can not fully account for the lensing is low effect, DESI shows a strong degeneracy along the $w_0$-$w_a$ axis, so that stronger deviations from {\lcdm} are definitely plausible, and both the lensing is low effect and the $S_8$ tension can potentially be alleviated by allowing for dynamic dark energy. Further studies combining DESI's BAO measurements with studies of Galaxy-Galaxy lensing and cosmic shear will be able to provide powerful constraints on the $w_0$-$w_a$ plane, potentially breaking the degeneracy present in current BAO and RSD measurements.

        We further identify that the main impact of dynamic dark energy on lensing observables is not from a different growth factor of the power spectrum. Instead, the difference stems from changes in the lensing efficiency prefactors, $\Delta\Sigma$ and $q_\kappa$, which are sensitive to changes in the distance-redshift relation. This is important to keep in mind e.g.~when investigating Fig.~6 of \citep{arxiv:2411.12022}, which shows that the posteriors of $S_8$ do not change between {\LCDM} and {\wcdm} in the DESI full-shape analysis of the galaxy power spectrum. From this, one might be tempted to conclude that the $S_8$ tension can not be sourced from this difference. However, this neglects the differences of how the lensing observables relate to the matter powerspectrum outlined above. When these are taken into account, dynamic dark energy with $w_0>-1$ and $w_a<0$ is a promising model to resolve the lensing is low effect and the $S_8$ tension.

        The latter insight is particularly crucial when one wants to analyze tensions in structure growth measurements in the presence of dynamic dark energy. Fig.~1 of \citet{2025PhRvD.111j3540S} compares changes in growth factor and distances for the \citet{2020A&A...641A...6P} best-fit {\lcdm} cosmology with both the DESI best-fit {\lcdm} cosmology and the DESI BAO cosmology. One can see that the differences in growth are highly degenerate, whereas the functional form of the changes to distances (and hence lensing efficiencies) differs. Considering that this change dominates at low redshifts, where direct distance measurements via BAOs at $\sim 100\,\mathrm{Mpc}/h$ are limited by a small volume, \textbf{a combined analysis of lensing and clustering offers the best way to determine whether growth tensions can be attributed to a departure from $\Lambda$.}

    \subsection{Caveats}
        This work constitutes an exploratory investigation of the potential of a $w_0w_a$CDM cosmology to resolve the lensing is low effect, the $S_8$ tension, and the evidence for strong baryon feedback. We do not claim to have performed a full analysis of these two effects, but rather to investigate if dynamical DE has the potential to "solve" these three cases in one fell swoop. We find that this is indeed the case for the lensing is low effect and the $S_8$ tension, and we find some interesting trends regarding baryon feedback, but we also note that there are several caveats to this work:
        \begin{itemize}
            \item We do not model the kinematic Sunyaev Zeldovich effect. The tSZ ($\propto M^{5/3}$) has a different mass dependence than kSZ ($\propto M$). Impacts on kSZ are thus so far hard to predict.
            \item Deviations from {\lcdm} are coupled to shifts in other cosmological parameters in the DESI results. Comparisons to a {\lcdm} cosmology with the same parameters is thus not necessarily a fair representation of a true cosmological parameter inference. We refer the reader again to Fig.~1 of \citet{2025PhRvD.111j3540S} to see how departures from $\Lambda$ impact growth and distance compared to changes in cosmlogical parameters within {\LCDM}.
        \end{itemize}
        While further work is certainly needed, this exploratory study serves to further motivate a rigorous investigation of dynamic dark energy models using a combination of lensing and clustering data.

\section{Conclusion}
\label{sec:conclusion}
        In this exploratory study, we find that a dynamic dark energy cosmology of the kind that recent DESI results favor ($w_0>-1,\,w_a<0$) has interesting impacts on lensing tensions. The combination of an increased (normalized) growth factor and decreased expansion rate at low redshift impacts observables differently. Lensing, which is sensitive to a combination of both growth and expansion, tends to be suppressed with respect to clustering observables. This alleviates both the lensing is low effect and the $S_8$ tension, but is not strong enough on its own to completely resolve them at DESI's best-fit cosmology. We note that the degeneracy along the $w_0$-$w_a$ plane are quite extended and stronger deviations from $\Lambda$ are possible, which would allow for a full alleviation of these tensions. We also note that its unique dependency on both expansion and growth makes lensing a powerful probe to detect deviations from {\lcdm} when combined with direct growth measurements (such as redshift-space distortions). This is particularly relevant at low redshifts, where a small volume leads to a strong sample variance in BAO measurements.
\section*{Acknowledgements}
\it
We are thankful to Chris Blake for making his covariance calculations available to us. SH is supported by the U.D Department of Energy, Office of Science, Office of High Energy Physics under Award Number DE-SC0019301.
\rm
\bibliographystyle{aa}
\bibliography{cite}

\begin{thebibliography}{57}
\expandafter\ifx\csname natexlab\endcsname\relax\def\natexlab#1{#1}\fi

\bibitem[{{Abdalla} {et~al.}(2022){Abdalla}, {Abell{\'a}n}, {Aboubrahim}, {Agnello}, {Akarsu}, {Akrami}, {Alestas}, {Aloni}, {Amendola}, {Anchordoqui}, {Anderson}, {Arendse}, {Asgari}, {Ballardini}, {Barger}, {Basilakos}, {Batista}, {Battistelli}, {Battye}, {Benetti}, {Benisty}, {Berlin}, {de Bernardis}, {Berti}, {Bidenko}, {Birrer}, {Blakeslee}, {Boddy}, {Bom}, {Bonilla}, {Borghi}, {Bouchet}, {Braglia}, {Buchert}, {Buckley-Geer}, {Calabrese}, {Caldwell}, {Camarena}, {Capozziello}, {Casertano}, {Chen}, {Chluba}, {Chen}, {Chen}, {Chudaykin}, {Cicoli}, {Copi}, {Courbin}, {Cyr-Racine}, {Czerny}, {Dainotti}, {D'Amico}, {Davis}, {de Cruz P{\'e}rez}, {de Haro}, {Delabrouille}, {Denton}, {Dhawan}, {Dienes}, {Di Valentino}, {Du}, {Eckert}, {Escamilla-Rivera}, {Fert{\'e}}, {Finelli}, {Fosalba}, {Freedman}, {Frusciante}, {Gazta{\~n}aga}, {Giar{\`e}}, {Giusarma}, {G{\'o}mez-Valent}, {Handley}, {Harrison}, {Hart}, {Hazra}, {Heavens}, {Heinesen}, {Hildebrandt}, {Hill}, {Hogg}, {Holz}, {Hooper}, {Hosseininejad}, {Huterer}, {Ishak}, {Ivanov}, {Jaffe}, {Jang}, {Jedamzik}, {Jimenez}, {Joseph}, {Joudaki}, {Kamionkowski}, {Karwal}, {Kazantzidis}, {Keeley}, {Klasen}, {Komatsu}, {Koopmans}, {Kumar}, {Lamagna}, {Lazkoz}, {Lee}, {Lesgourgues}, {Levi Said}, {Lewis}, {L'Huillier}, {Lucca}, {Maartens}, {Macri}, {Marfatia}, {Marra}, {Martins}, {Masi}, {Matarrese}, {Mazumdar}, {Melchiorri}, {Mena}, {Mersini-Houghton}, {Mertens}, {Milakovi{\'c}}, {Minami}, {Miranda}, {Moreno-Pulido}, {Moresco}, {Mota}, {Mottola}, {Mozzon}, {Muir}, {Mukherjee}, {Mukherjee}, {Naselsky}, {Nath}, {Nesseris}, {Niedermann}, {Notari}, {Nunes}, {{\'O} Colg{\'a}in}, {Owens}, {{\"O}z{\"u}lker}, {Pace}, {Paliathanasis}, {Palmese}, {Pan}, {Paoletti}, {Perez Bergliaffa}, {Perivolaropoulos}, {Pesce}, {Pettorino}, {Philcox}, {Pogosian}, {Poulin}, {Poulot}, {Raveri}, {Reid}, {Renzi}, {Riess}, {Sabla}, {Salucci}, {Salzano}, {Saridakis}, {Sathyaprakash}, {Schmaltz}, {Sch{\"o}neberg}, {Scolnic}, {Sen}, {Sehgal}, {Shafieloo}, {Sheikh-Jabbari}, {Silk}, {Silvestri}, {Skara}, {Sloth}, {Soares-Santos}, {Sol{\`a} Peracaula}, {Songsheng}, {Soriano}, {Staicova}, {Starkman}, {Szapudi}, {Teixeira}, {Thomas}, {Treu}, {Trott}, {van de Bruck}, {Vazquez}, {Verde}, {Visinelli}, {Wang}, {Wang}, {Wang}, {Watkins}, {Watson}, {Webb}, {Weiner}, {Weltman}, {Witte}, {Wojtak}, \& {Yadav}}]{2022JHEAp..34...49A}
{Abdalla}, E., {Abell{\'a}n}, G.~F., {Aboubrahim}, A., {et~al.} 2022, Journal of High Energy Astrophysics, 34, 49

\bibitem[{{Adame} {et~al.}(2025{\natexlab{a}}){Adame}, {Aguilar}, {Ahlen}, {Alam}, {Alexander}, {Allende Prieto}, {Alvarez}, {Alves}, {Anand}, {Andrade}, {Armengaud}, {Avila}, {Aviles}, {Awan}, {Bahr-Kalus}, {Bailey}, {Baltay}, {Bault}, {Behera}, {BenZvi}, {Beutler}, {Bianchi}, {Blake}, {Blum}, {Bonici}, {Brieden}, {Brodzeller}, {Brooks}, {Buckley-Geer}, {Burtin}, {Calderon}, {Canning}, {Carnero Rosell}, {Cereskaite}, {Cervantes-Cota}, {Chabanier}, {Chaussidon}, {Chaves-Montero}, {Chebat}, {Chen}, {Chen}, {Claybaugh}, {Cole}, {Cuceu}, {Davis}, {Dawson}, {de la Macorra}, {de Mattia}, {Deiosso}, {Dey}, {Dey}, {Ding}, {Doel}, {Edelstein}, {Eftekharzadeh}, {Eisenstein}, {Elbers}, {Elliott}, {Fagrelius}, {Fanning}, {Ferraro}, {Ereza}, {Findlay}, {Flaugher}, {Font-Ribera}, {Forero-S{\'a}nchez}, {Forero-Romero}, {Frenk}, {Garcia-Quintero}, {Garrison}, {Gazta{\~n}aga}, {Gil-Mar{\'\i}n}, {Gontcho}, {Gonzalez-Morales}, {Gonzalez-Perez}, {Gordon}, {Green}, {Gruen}, {Gsponer}, {Gutierrez}, {Guy}, {Hadzhiyska}, {Hahn}, {Hanif}, {Herrera-Alcantar}, {Honscheid}, {Howlett}, {Huterer}, {Ir{\v{s}}i{\v{c}}}, {Ishak}, {Joyce}, {Juneau}, {Kara{\c{c}}ayl{\i}}, {Kehoe}, {Kent}, {Kirkby}, {Kong}, {Koposov}, {Kremin}, {Krolewski}, {Lahav}, {Lai}, {Lan}, {Landriau}, {Lang}, {Lasker}, {Le Goff}, {Le Guillou}, {Leauthaud}, {Levi}, {Li}, {Lodha}, {Magneville}, {Manera}, {Margala}, {Martini}, {Matthewson}, {Maus}, {McDonald}, {Medina-Varela}, {Meisner}, {Mena-Fern{\'a}ndez}, {Miquel}, {Moon}, {Moore}, {Moustakas}, {Mudur}, {Mueller}, {Mu{\~n}oz-Guti{\'e}rrez}, {Myers}, {Nadathur}, {Napolitano}, {Neveux}, {Newman}, {Nguyen}, {Nie}, {Niz}, {Noriega}, {Padmanabhan}, {Paillas}, {Palanque-Delabrouille}, {Pan}, {Penmetsa}, {Percival}, {Pieri}, {Pinon}, {Poppett}, {Porredon}, {Prada}, {P{\'e}rez-Fern{\'a}ndez}, {P{\'e}rez-R{\`a}fols}, {Rabinowitz}, {Raichoor}, {Ram{\'\i}rez-P{\'e}rez}, {Ramirez-Solano}, {Rashkovetskyi}, {Ravoux}, {Rezaie}, {Rich}, {Rocher}, {Rockosi}, {Roe}, {Rosado-Marin}, {Ross}, {Rossi}, {Ruggeri}, {Ruhlmann-Kleider}, {Samushia}, {Sanchez}, {Saulder}, {Schlafly}, {Schlegel}, {Schubnell}, {Seo}, {Shafieloo}, {Sharples}, {Silber}, {Slosar}, {Smith}, {Sprayberry}, {Tan}, {Tarl{\'e}}, {Taylor}, {Trusov}, {Vaisakh}, {Valcin}, {Valdes}, {Valogiannis}, {Vargas-Maga{\~n}a}, {Verde}, {Walther}, {Wang}, {Wang}, {Weaver}, {Weaverdyck}, {Wechsler}, {Weinberg}, {White}, {Wilson}, \& {Yi}}]{arxiv:2411.12022}
{Adame}, A.~G., {Aguilar}, J., {Ahlen}, S., {et~al.} 2025{\natexlab{a}}, \jcap, 2025, 028

\bibitem[{{Adame} {et~al.}(2025{\natexlab{b}}){Adame}, {Aguilar}, {Ahlen}, {Alam}, {Alexander}, {Alvarez}, {Alves}, {Anand}, {Andrade}, {Armengaud}, {Avila}, {Aviles}, {Awan}, {Bahr-Kalus}, {Bailey}, {Baltay}, {Bault}, {Behera}, {BenZvi}, {Bera}, {Beutler}, {Bianchi}, {Blake}, {Blum}, {Brieden}, {Brodzeller}, {Brooks}, {Buckley-Geer}, {Burtin}, {Calderon}, {Canning}, {Carnero Rosell}, {Cereskaite}, {Cervantes-Cota}, {Chabanier}, {Chaussidon}, {Chaves-Montero}, {Chen}, {Chen}, {Claybaugh}, {Cole}, {Cuceu}, {Davis}, {Dawson}, {de la Macorra}, {de Mattia}, {Deiosso}, {Dey}, {Dey}, {Ding}, {Doel}, {Edelstein}, {Eftekharzadeh}, {Eisenstein}, {Elliott}, {Fagrelius}, {Fanning}, {Ferraro}, {Ereza}, {Findlay}, {Flaugher}, {Font-Ribera}, {Forero-S{\'a}nchez}, {Forero-Romero}, {Frenk}, {Garcia-Quintero}, {Gazta{\~n}aga}, {Gil-Mar{\'\i}n}, {Gontcho a Gontcho}, {Gonzalez-Morales}, {Gonzalez-Perez}, {Gordon}, {Green}, {Gruen}, {Gsponer}, {Gutierrez}, {Guy}, {Hadzhiyska}, {Hahn}, {Hanif}, {Herrera-Alcantar}, {Honscheid}, {Howlett}, {Huterer}, {Ir{\v{s}}i{\v{c}}}, {Ishak}, {Juneau}, {Kara{\c{c}}ayl{\i}}, {Kehoe}, {Kent}, {Kirkby}, {Kremin}, {Krolewski}, {Lai}, {Lan}, {Landriau}, {Lang}, {Lasker}, {Le Goff}, {Le Guillou}, {Leauthaud}, {Levi}, {Li}, {Linder}, {Lodha}, {Magneville}, {Manera}, {Margala}, {Martini}, {Maus}, {McDonald}, {Medina-Varela}, {Meisner}, {Mena-Fern{\'a}ndez}, {Miquel}, {Moon}, {Moore}, {Moustakas}, {Mueller}, {Mu{\~n}oz-Guti{\'e}rrez}, {Myers}, {Nadathur}, {Napolitano}, {Neveux}, {Newman}, {Nguyen}, {Nie}, {Niz}, {Noriega}, {Padmanabhan}, {Paillas}, {Palanque-Delabrouille}, {Pan}, {Penmetsa}, {Percival}, {Pieri}, {Pinon}, {Poppett}, {Porredon}, {Prada}, {P{\'e}rez-Fern{\'a}ndez}, {P{\'e}rez-R{\`a}fols}, {Rabinowitz}, {Raichoor}, {Ram{\'\i}rez-P{\'e}rez}, {Ramirez-Solano}, {Rashkovetskyi}, {Ravoux}, {Rezaie}, {Rich}, {Rocher}, {Rockosi}, {Roe}, {Rosado-Marin}, {Ross}, {Rossi}, {Ruggeri}, {Ruhlmann-Kleider}, {Samushia}, {Sanchez}, {Saulder}, {Schlafly}, {Schlegel}, {Schubnell}, {Seo}, {Shafieloo}, {Sharples}, {Silber}, {Slosar}, {Smith}, {Sprayberry}, {Tan}, {Tarl{\'e}}, {Taylor}, {Trusov}, {Ure{\~n}a-L{\'o}pez}, {Vaisakh}, {Valcin}, {Valdes}, {Vargas-Maga{\~n}a}, {Verde}, {Walther}, {Wang}, {Wang}, {Weaver}, {Weaverdyck}, {Wechsler}, {Weinberg}, {White}, {Yu}, {Yu}, {Yuan}, {Y{\`e}che}, {Zaborowski}, {Zarrouk}, {Zhang}, {Zhao}, {Zhao}, {Zhou}, \& {Zhuang}}]{2025JCAP...02..021A}
{Adame}, A.~G., {Aguilar}, J., {Ahlen}, S., {et~al.} 2025{\natexlab{b}}, \jcap, 2025, 021

\bibitem[{{Amon} {et~al.}(2022){Amon}, {Gruen}, {Troxel}, {MacCrann}, {Dodelson}, {Choi}, {Doux}, {Secco}, {Samuroff}, {Krause}, {Cordero}, {Myles}, {DeRose}, {Wechsler}, {Gatti}, {Navarro-Alsina}, {Bernstein}, {Jain}, {Blazek}, {Alarcon}, {Fert{\'e}}, {Lemos}, {Raveri}, {Campos}, {Prat}, {S{\'a}nchez}, {Jarvis}, {Alves}, {Andrade-Oliveira}, {Baxter}, {Bechtol}, {Becker}, {Bridle}, {Camacho}, {Carnero Rosell}, {Carrasco Kind}, {Cawthon}, {Chang}, {Chen}, {Chintalapati}, {Crocce}, {Davis}, {Diehl}, {Drlica-Wagner}, {Eckert}, {Eifler}, {Elvin-Poole}, {Everett}, {Fang}, {Fosalba}, {Friedrich}, {Gaztanaga}, {Giannini}, {Gruendl}, {Harrison}, {Hartley}, {Herner}, {Huang}, {Huff}, {Huterer}, {Kuropatkin}, {Leget}, {Liddle}, {McCullough}, {Muir}, {Pandey}, {Park}, {Porredon}, {Refregier}, {Rollins}, {Roodman}, {Rosenfeld}, {Ross}, {Rykoff}, {Sanchez}, {Sevilla-Noarbe}, {Sheldon}, {Shin}, {Troja}, {Tutusaus}, {Tutusaus}, {Varga}, {Weaverdyck}, {Yanny}, {Yin}, {Zhang}, {Zuntz}, {Aguena}, {Allam}, {Annis}, {Bacon}, {Bertin}, {Bhargava}, {Brooks}, {Buckley-Geer}, {Burke}, {Carretero}, {Costanzi}, {da Costa}, {Pereira}, {De Vicente}, {Desai}, {Dietrich}, {Doel}, {Ferrero}, {Flaugher}, {Frieman}, {Garc{\'\i}a-Bellido}, {Gaztanaga}, {Gerdes}, {Giannantonio}, {Gschwend}, {Gutierrez}, {Hinton}, {Hollowood}, {Honscheid}, {Hoyle}, {James}, {Kron}, {Kuehn}, {Lahav}, {Lima}, {Lin}, {Maia}, {Marshall}, {Martini}, {Melchior}, {Menanteau}, {Miquel}, {Mohr}, {Morgan}, {Ogando}, {Palmese}, {Paz-Chinch{\'o}n}, {Petravick}, {Pieres}, {Romer}, {Sanchez}, {Scarpine}, {Schubnell}, {Serrano}, {Smith}, {Soares-Santos}, {Tarle}, {Thomas}, {To}, {Weller}, \& {DES Collaboration}}]{Amon:2022}
{Amon}, A., {Gruen}, D., {Troxel}, M.~A., {et~al.} 2022, \prd, 105, 023514

\bibitem[{{Bartelmann} \& {Schneider}(2001)}]{2001PhR...340..291B}
{Bartelmann}, M. \& {Schneider}, P. 2001, \physrep, 340, 291

\bibitem[{{Bigwood} {et~al.}(2024){Bigwood}, {Amon}, {Schneider}, {Salcido}, {McCarthy}, {Preston}, {Sanchez}, {Sijacki}, {Schaan}, {Ferraro}, {Battaglia}, {Chen}, {Dodelson}, {Roodman}, {Pieres}, {Fert{\'e}}, {Alarcon}, {Drlica-Wagner}, {Choi}, {Navarro-Alsina}, {Campos}, {Ross}, {Carnero Rosell}, {Yin}, {Yanny}, {S{\'a}nchez}, {Chang}, {Davis}, {Doux}, {Gruen}, {Rykoff}, {Huff}, {Sheldon}, {Tarsitano}, {Andrade-Oliveira}, {Bernstein}, {Giannini}, {Diehl}, {Huang}, {Harrison}, {Sevilla-Noarbe}, {Tutusaus}, {Elvin-Poole}, {McCullough}, {Zuntz}, {Blazek}, {DeRose}, {Cordero}, {Prat}, {Myles}, {Eckert}, {Bechtol}, {Herner}, {Secco}, {Gatti}, {Raveri}, {Kind}, {Becker}, {Troxel}, {Jarvis}, {MacCrann}, {Friedrich}, {Alves}, {Leget}, {Chen}, {Rollins}, {Wechsler}, {Gruendl}, {Cawthon}, {Allam}, {Bridle}, {Pandey}, {Everett}, {Shin}, {Hartley}, {Fang}, {Zhang}, {Aguena}, {Annis}, {Bacon}, {Bertin}, {Bocquet}, {Brooks}, {Carretero}, {Castander}, {da Costa}, {Pereira}, {De Vicente}, {Desai}, {Doel}, {Ferrero}, {Flaugher}, {Frieman}, {Garc{\'\i}a-Bellido}, {Gaztanaga}, {Gutierrez}, {Hinton}, {Hollowood}, {Honscheid}, {Huterer}, {James}, {Kuehn}, {Lahav}, {Lee}, {Marshall}, {Mena-Fern{\'a}ndez}, {Miquel}, {Muir}, {Paterno}, {Plazas Malag{\'o}n}, {Porredon}, {Romer}, {Samuroff}, {Sanchez}, {Sanchez Cid}, {Smith}, {Soares-Santos}, {Suchyta}, {Swanson}, {Tarle}, {To}, {Weaverdyck}, {Weller}, {Wiseman}, \& {Yamamoto}}]{Bigwood:2024}
{Bigwood}, L., {Amon}, A., {Schneider}, A., {et~al.} 2024, \mnras, 534, 655

\bibitem[{{Chaves-Montero} {et~al.}(2023){Chaves-Montero}, {Angulo}, \& {Contreras}}]{arxiv:2211.01744}
{Chaves-Montero}, J., {Angulo}, R.~E., \& {Contreras}, S. 2023, \mnras, 521, 937

\bibitem[{{Chen} {et~al.}(2023){Chen}, {Aric{\`o}}, {Huterer}, {Angulo}, {Weaverdyck}, {Friedrich}, {Secco}, {Hern{\'a}ndez-Monteagudo}, {Alarcon}, {Alves}, {Amon}, {Andrade-Oliveira}, {Baxter}, {Bechtol}, {Becker}, {Bernstein}, {Blazek}, {Brandao-Souza}, {Bridle}, {Camacho}, {Campos}, {Carnero Rosell}, {Carrasco Kind}, {Cawthon}, {Chang}, {Chen}, {Chintalapati}, {Choi}, {Cordero}, {Crocce}, {Pereira}, {Davis}, {DeRose}, {Di Valentino}, {Diehl}, {Dodelson}, {Doux}, {Drlica-Wagner}, {Eckert}, {Eifler}, {Elsner}, {Elvin-Poole}, {Everett}, {Fang}, {Fert{\'e}}, {Fosalba}, {Gatti}, {Gaztanaga}, {Giannini}, {Gruen}, {Gruendl}, {Harrison}, {Hartley}, {Herner}, {Hoffmann}, {Huang}, {Huff}, {Jain}, {Jarvis}, {Jeffrey}, {Kacprzak}, {Krause}, {Kuropatkin}, {Leget}, {Lemos}, {Liddle}, {MacCrann}, {McCullough}, {Muir}, {Myles}, {Navarro-Alsina}, {Omori}, {Pandey}, {Park}, {Porredon}, {Prat}, {Raveri}, {Refregier}, {Rollins}, {Roodman}, {Rosenfeld}, {Ross}, {Rykoff}, {Samuroff}, {S{\'a}nchez}, {Sanchez}, {Sevilla-Noarbe}, {Sheldon}, {Shin}, {Troja}, {Troxel}, {Tutusaus}, {Varga}, {Wechsler}, {Yanny}, {Yin}, {Zhang}, {Zuntz}, {Aguena}, {Annis}, {Bacon}, {Bertin}, {Bocquet}, {Brooks}, {Burke}, {Carretero}, {Conselice}, {Costanzi}, {da Costa}, {De Vicente}, {Desai}, {Doel}, {Ferrero}, {Flaugher}, {Frieman}, {Garc{\'\i}a-Bellido}, {Gerdes}, {Giannantonio}, {Gschwend}, {Gutierrez}, {Hinton}, {Hollowood}, {Honscheid}, {James}, {Kuehn}, {Lahav}, {March}, {Marshall}, {Melchior}, {Menanteau}, {Miquel}, {Mohr}, {Morgan}, {Paz-Chinch{\'o}n}, {Pieres}, {Sanchez}, {Smith}, {Suchyta}, {Swanson}, {Tarle}, {Thomas}, {To}, \& {DES Collaboration}}]{arxiv:2206.08591}
{Chen}, A., {Aric{\`o}}, G., {Huterer}, D., {et~al.} 2023, \mnras, 518, 5340

\bibitem[{{Chen} {et~al.}(2024){Chen}, {DeRose}, {Zhou}, {White}, {Ferraro}, {Blake}, {Lange}, {Wechsler}, {Aguilar}, {Ahlen}, {Brooks}, {Claybaugh}, {Dawson}, {de la Macorra}, {Doel}, {Font-Ribera}, {Gazta{\~n}aga}, {Gontcho A Gontcho}, {Gutierrez}, {Honscheid}, {Howlett}, {Kehoe}, {Kirkby}, {Kisner}, {Kremin}, {Landriau}, {Le Guillou}, {Manera}, {Meisner}, {Miquel}, {Newman}, {Niz}, {Palanque-Delabrouille}, {Percival}, {Prada}, {Rossi}, {Sanchez}, {Schlegel}, {Schubnell}, {Sprayberry}, {Tarl{\'e}}, \& {Weaver}}]{arxiv:2407.04795}
{Chen}, S., {DeRose}, J., {Zhou}, R., {et~al.} 2024, \prd, 110, 103518

\bibitem[{{Chevallier} \& {Polarski}(2001)}]{arXiv:gr-qc/0009008}
{Chevallier}, M. \& {Polarski}, D. 2001, International Journal of Modern Physics D, 10, 213

\bibitem[{{Chisari} {et~al.}(2019){Chisari}, {Alonso}, {Krause}, {Leonard}, {Bull}, {Neveu}, {Villarreal}, {Singh}, {McClintock}, {Ellison}, {Du}, {Zuntz}, {Mead}, {Joudaki}, {Lorenz}, {Tr{\"o}ster}, {Sanchez}, {Lanusse}, {Ishak}, {Hlozek}, {Blazek}, {Campagne}, {Almoubayyed}, {Eifler}, {Kirby}, {Kirkby}, {Plaszczynski}, {Slosar}, {Vrastil}, {Wagoner}, \& {LSST Dark Energy Science Collaboration}}]{arXiv:1812.05995}
{Chisari}, N.~E., {Alonso}, D., {Krause}, E., {et~al.} 2019, \apjs, 242, 2

\bibitem[{{Contreras} {et~al.}(2023){Contreras}, {Chaves-Montero}, \& {Angulo}}]{2023MNRAS.525.3149C}
{Contreras}, S., {Chaves-Montero}, J., \& {Angulo}, R.~E. 2023, \mnras, 525, 3149

\bibitem[{{Cooray} \& {Sheth}(2002)}]{2002PhR...372....1C}
{Cooray}, A. \& {Sheth}, R. 2002, \physrep, 372, 1

\bibitem[{{Dark Energy Survey and Kilo-Degree Survey Collaboration} {et~al.}(2023){Dark Energy Survey and Kilo-Degree Survey Collaboration}, {Abbott}, {Aguena}, {Alarcon}, {Alves}, {Amon}, {Andrade-Oliveira}, {Asgari}, {Avila}, {Bacon}, {Bechtol}, {Becker}, {Bernstein}, {Bertin}, {Bilicki}, {Blazek}, {Bocquet}, {Brooks}, {Burger}, {Burke}, {Camacho}, {Campos}, {Carnero Rosell}, {Carrasco Kind}, {Carretero}, {Castander}, {Cawthon}, {Chang}, {Chen}, {Choi}, {Conselice}, {Cordero}, {Crocce}, {da Costa}, {da Silva Pereira}, {Dalal}, {Davis}, {de Jong}, {DeRose}, {Desai}, {Diehl}, {Dodelson}, {Doel}, {Doux}, {Drlica-Wagner}, {Dvornik}, {Eckert}, {Eifler}, {Elvin-Poole}, {Everett}, {Fang}, {Ferrero}, {Fert{\'e}}, {Flaugher}, {Friedrich}, {Frieman}, {Garc{\'\i}a-Bellido}, {Gatti}, {Giannini}, {Giblin}, {Gruen}, {Gruendl}, {Gutierrez}, {Harrison}, {Hartley}, {Herner}, {Heymans}, {Hildebrandt}, {Hinton}, {Hoekstra}, {Hollowood}, {Honscheid}, {Huang}, {Huff}, {Huterer}, {James}, {Jarvis}, {Jeffrey}, {Jeltema}, {Joachimi}, {Joudaki}, {Kannawadi}, {Krause}, {Kuehn}, {Kuijken}, {Kuropatkin}, {Lahav}, {Leget}, {Lemos}, {Li}, {Li}, {Liddle}, {Lima}, {Lin}, {Lin}, {MacCrann}, {Mahony}, {Marshall}, {McCullough}, {Mena-Fern{\'a}ndez}, {Menanteau}, {Miquel}, {Mohr}, {Muir}, {Myles}, {Napolitano}, {Navarro-Alsina}, {Ogando}, {Palmese}, {Pandey}, {Park}, {Paterno}, {Peacock}, {Petravick}, {Pieres}, {Plazas Malag{\'o}n}, {Porredon}, {Prat}, {Radovich}, {Raveri}, {Reischke}, {Robertson}, {Rollins}, {Romer}, {Roodman}, {Rykoff}, {Samuroff}, {S{\'a}nchez}, {Sanchez}, {Sanchez}, {Schneider}, {Secco}, {Sevilla-Noarbe}, {Shan}, {Sheldon}, {Shin}, {Sif{\'o}n}, {Smith}, {Soares-Santos}, {St{\"o}lzner}, {Suchyta}, {Swanson}, {Tarle}, {Thomas}, {To}, {Troxel}, {Tr{\"o}ster}, {Tutusaus}, {van den Busch}, {Varga}, {Walker}, {Weaverdyck}, {Wechsler}, {Weller}, {Wiseman}, {Wright}, {Yanny}, {Yin}, {Yoon}, {Zhang}, \& {Zuntz}}]{2023OJAp....6E..36D}
{Dark Energy Survey and Kilo-Degree Survey Collaboration}, {Abbott}, T.~M.~C., {Aguena}, M., {et~al.} 2023, The Open Journal of Astrophysics, 6, 36

\bibitem[{{DESI Collaboration} {et~al.}(2025){DESI Collaboration}, {Abdul-Karim}, {Aguilar}, {Ahlen}, {Alam}, {Allen}, {Allende Prieto}, {Alves}, {Anand}, {Andrade}, {Armengaud}, {Aviles}, {Bailey}, {Baltay}, {Bansal}, {Bault}, {Behera}, {BenZvi}, {Bianchi}, {Blake}, {Brieden}, {Brodzeller}, {Brooks}, {Buckley-Geer}, {Burtin}, {Calderon}, {Canning}, {Carnero Rosell}, {Carrilho}, {Casas}, {Castander}, {Cereskaite}, {Charles}, {Chaussidon}, {Chaves-Montero}, {Chebat}, {Chen}, {Claybaugh}, {Cole}, {Cooper}, {Cuceu}, {Dawson}, {de la Macorra}, {de Mattia}, {Deiosso}, {Della Costa}, {Demina}, {Dey}, {Dey}, {Ding}, {Doel}, {Edelstein}, {Eisenstein}, {Elbers}, {Fagrelius}, {Fanning}, {Fern{\'a}ndez-Garc{\'i}a}, {Ferraro}, {Font-Ribera}, {Forero-Romero}, {Frenk}, {Garcia-Quintero}, {Garrison}, {Gazta{\~n}aga}, {Gil-Mar{\'i}n}, {Gontcho}, {Gonzalez}, {Gonzalez-Morales}, {Gordon}, {Green}, {Gutierrez}, {Guy}, {Hadzhiyska}, {Hahn}, {He}, {Herbold}, {Herrera-Alcantar}, {Ho}, {Honscheid}, {Howlett}, {Huterer}, {Ishak}, {Juneau}, {Kamble}, {Kara{\c c}ayl{\i}}, {Kehoe}, {Kent}, {Kim}, {Kirkby}, {Kisner}, {Koposov}, {Kremin}, {Krolewski}, {Lahav}, {Lamman}, {Landriau}, {Lang}, {Lasker}, {Le Goff}, {Le Guillou}, {Leauthaud}, {Levi}, {Li}, {Li}, {Lodha}, {Lokken}, {Lozano-Rodr{\'i}guez}, {Magneville}, {Manera}, {Martini}, {Matthewson}, {Meisner}, {Mena-Fern{\'a}ndez}, {Menegas}, {Mergulh{\~a}o}, {Miquel}, {Moustakas}, {Mu{\~n}oz-Guti{\'e}rrez}, {Mu{\~n}oz-Santos}, {Myers}, {Nadathur}, {Naidoo}, {Napolitano}, {Newman}, {Niz}, {Noriega}, {Paillas}, {Palanque-Delabrouille}, {Pan}, {Peacock}, {Pellejero Ibanez}, {Percival}, {P{\'e}rez-Fern{\'a}ndez}, {P{\'e}rez-R{\`a}fols}, {Pieri}, {Poppett}, {Prada}, {Rabinowitz}, {Raichoor}, {Ram{\'i}rez-P{\'e}rez}, {Rashkovetskyi}, {Ravoux}, {Rich}, {Rocher}, {Rockosi}, {Rohlf}, {Rom{\'a}n-Herrera}, {Ross}, {Rossi}, {Ruggeri}, {Ruhlmann-Kleider}, {Samushia}, {Sanchez}, {Sanders}, {Schlegel}, {Schubnell}, {Seo}, {Shafieloo}, {Sharples}, {Silber}, {Sinigaglia}, {Sprayberry}, {Tan}, {Tarl{\'e}}, {Taylor}, {Turner}, {Ure{\~n}a-L{\'o}pez}, {Vaisakh}, {Valdes}, {Valogiannis}, {Vargas-Maga{\~n}a}, {Verde}, {Walther}, {Weaver}, {Weinberg}, {White}, {Wolfson}, {Y{\`e}che}, {Yu}, {Zaborowski}, {Zarrouk}, {Zhai}, {Zhang}, {Zhao}, {Zhao}, {Zhou}, \& {Zou}}]{arXiv:2503.14738}
{DESI Collaboration}, {Abdul-Karim}, M., {Aguilar}, J., {et~al.} 2025, arXiv e-prints, arXiv:2503.14738

\bibitem[{{Eisenstein} \& {Hu}(1998)}]{1998ApJ...496..605E}
{Eisenstein}, D.~J. \& {Hu}, W. 1998, \apj, 496, 605

\bibitem[{{Elbers} {et~al.}(2025){Elbers}, {Aviles}, {Noriega}, {Chebat}, {Menegas}, {Frenk}, {Garcia-Quintero}, {Gonzalez}, {Ishak}, {Lahav}, {Naidoo}, {Niz}, {Y{\`e}che}, {Abdul-Karim}, {Ahlen}, {Alves}, {Andrade}, {Armengaud}, {BenZvi}, {Bianchi}, {Brieden}, {Brodzeller}, {Brooks}, {Burtin}, {Calderon}, {Canning}, {Carnero Rosell}, {Casas}, {Castander}, {Charles}, {Chaussidon}, {Chaves-Montero}, {Claybaugh}, {Cole}, {Cooper}, {Cuceu}, {Dawson}, {de la Macorra}, {de Mattia}, {Deiosso}, {Dey}, {Dey}, {Ding}, {Doel}, {Eisenstein}, {Ferraro}, {Font-Ribera}, {Forero-Romero}, {Garrison}, {Gazta{\~n}aga}, {Gil-Mar{\'\i}n}, {Gontcho}, {Gonzalez-Morales}, {Gutierrez}, {He}, {Herbold}, {Herrera-Alcantar}, {Howlett}, {Huterer}, {Juneau}, {Kehoe}, {Kirkby}, {Kisner}, {Kremin}, {Lamman}, {Landriau}, {Le Guillou}, {Leauthaud}, {Levi}, {Li}, {Lodha}, {Magneville}, {Manera}, {Martini}, {Matthewson}, {Meisner}, {Mena-Fern{\'a}ndez}, {Miquel}, {Moustakas}, {Nadathur}, {Newman}, {Paillas}, {Palanque-Delabrouille}, {Percival}, {Pieri}, {Poppett}, {Prada}, {P{\'e}rez-R{\`a}fols}, {Rabinowitz}, {Ram{\'\i}rez-P{\'e}rez}, {Rashkovetskyi}, {Ravoux}, {Rivera-Morales}, {Rohlf}, {Ross}, {Rossi}, {Ruhlmann-Kleider}, {Samushia}, {Sanchez}, {Schlegel}, {Schubnell}, {Seo}, {Sinigaglia}, {Sprayberry}, {Tan}, {Tarl{\'e}}, {Taylor}, {Turner}, {Vargas-Maga{\~n}a}, {Verde}, {Walther}, {Weaver}, {Whitford}, {Wolfson}, {Y{\`e}che}, {Zarrouk}, {Zhao}, {Zhou}, \& {Zou}}]{arXiv:2503.14744}
{Elbers}, W., {Aviles}, A., {Noriega}, H.~E., {et~al.} 2025, arXiv e-prints, arXiv:2503.14744

\bibitem[{{Friedrich} {et~al.}(2021){Friedrich}, {Andrade-Oliveira}, {Camacho}, {Alves}, {Rosenfeld}, {Sanchez}, {Fang}, {Eifler}, {Krause}, {Chang}, {Omori}, {Amon}, {Baxter}, {Elvin-Poole}, {Huterer}, {Porredon}, {Prat}, {Terra}, {Troja}, {Alarcon}, {Bechtol}, {Bernstein}, {Buchs}, {Campos}, {Carnero Rosell}, {Carrasco Kind}, {Cawthon}, {Choi}, {Cordero}, {Crocce}, {Davis}, {DeRose}, {Diehl}, {Dodelson}, {Doux}, {Drlica-Wagner}, {Elsner}, {Everett}, {Fosalba}, {Gatti}, {Giannini}, {Gruen}, {Gruendl}, {Harrison}, {Hartley}, {Jain}, {Jarvis}, {MacCrann}, {McCullough}, {Muir}, {Myles}, {Pandey}, {Raveri}, {Roodman}, {Rodriguez-Monroy}, {Rykoff}, {Samuroff}, {S{\'a}nchez}, {Secco}, {Sevilla-Noarbe}, {Sheldon}, {Troxel}, {Weaverdyck}, {Yanny}, {Aguena}, {Avila}, {Bacon}, {Bertin}, {Bhargava}, {Brooks}, {Burke}, {Carretero}, {Costanzi}, {da Costa}, {Pereira}, {De Vicente}, {Desai}, {Evrard}, {Ferrero}, {Frieman}, {Garc{\'\i}a-Bellido}, {Gaztanaga}, {Gerdes}, {Giannantonio}, {Gschwend}, {Gutierrez}, {Hinton}, {Hollowood}, {Honscheid}, {James}, {Kuehn}, {Lahav}, {Lima}, {Maia}, {Menanteau}, {Miquel}, {Morgan}, {Palmese}, {Paz-Chinch{\'o}n}, {Plazas}, {Sanchez}, {Scarpine}, {Serrano}, {Soares-Santos}, {Smith}, {Suchyta}, {Tarle}, {Thomas}, {To}, {Varga}, {Weller}, {Wilkinson}, {Wilkinson}, \& {DES Collaboration}}]{arxiv:2012.08568}
{Friedrich}, O., {Andrade-Oliveira}, F., {Camacho}, H., {et~al.} 2021, \mnras, 508, 3125

\bibitem[{{Garrison} {et~al.}(2021){Garrison}, {Eisenstein}, {Ferrer}, {Maksimova}, \& {Pinto}}]{2021MNRAS.508..575G}
{Garrison}, L.~H., {Eisenstein}, D.~J., {Ferrer}, D., {Maksimova}, N.~A., \& {Pinto}, P.~A. 2021, \mnras, 508, 575

\bibitem[{{Greco} {et~al.}(2015){Greco}, {Hill}, {Spergel}, \& {Battaglia}}]{2015ApJ...808..151G}
{Greco}, J.~P., {Hill}, J.~C., {Spergel}, D.~N., \& {Battaglia}, N. 2015, \apj, 808, 151

\bibitem[{{Hadzhiyska} {et~al.}(2024){Hadzhiyska}, {Ferraro}, {Ried Guachalla}, {Schaan}, {Aguilar}, {Battaglia}, {Bond}, {Brooks}, {Calabrese}, {Choi}, {Claybaugh}, {Coulton}, {Dawson}, {Devlin}, {Dey}, {Doel}, {Duivenvoorden}, {Dunkley}, {Farren}, {Font-Ribera}, {Forero-Romero}, {Gallardo}, {Gazta{\~n}aga}, {Gontcho Gontcho}, {Gralla}, {Le Guillou}, {Gutierrez}, {Guy}, {Hill}, {Hlo{\v{z}}ek}, {Honscheid}, {Juneau}, {Kisner}, {Kremin}, {Landriau}, {Liu}, {Louis}, {MacCrann}, {de Macorra}, {Madhavacheril}, {Manera}, {Meisner}, {Miquel}, {Moodley}, {Moustakas}, {Mroczkowski}, {Naess}, {Newman}, {Niemack}, {Niz}, {Page}, {Palanque-Delabrouille}, {Partridge}, {Percival}, {Prada}, {Qu}, {Rossi}, {Sanchez}, {Schlegel}, {Schubnell}, {Sehgal}, {Seo}, {Sif{\'o}n}, {Spergel}, {Sprayberry}, {Staggs}, {Tarl{\'e}}, {Vargas}, {Vavagiakis}, {Weaver}, {Wollack}, {Zhou}, \& {Zou}}]{Hadzhiyska:2024}
{Hadzhiyska}, B., {Ferraro}, S., {Ried Guachalla}, B., {et~al.} 2024, arXiv e-prints, arXiv:2407.07152

\bibitem[{{Hahn} {et~al.}(2023){Hahn}, {Wilson}, {Ruiz-Macias}, {Cole}, {Weinberg}, {Moustakas}, {Kremin}, {Tinker}, {Smith}, {Wechsler}, {Ahlen}, {Alam}, {Bailey}, {Brooks}, {Cooper}, {Davis}, {Dawson}, {Dey}, {Dey}, {Eftekharzadeh}, {Eisenstein}, {Fanning}, {Forero-Romero}, {Frenk}, {Gazta{\~n}aga}, {A Gontcho}, {Guy}, {Honscheid}, {Ishak}, {Juneau}, {Kehoe}, {Kisner}, {Lan}, {Landriau}, {Le Guillou}, {Levi}, {Magneville}, {Martini}, {Meisner}, {Myers}, {Nie}, {Norberg}, {Palanque-Delabrouille}, {Percival}, {Poppett}, {Prada}, {Raichoor}, {Ross}, {Gaines}, {Saulder}, {Schlafly}, {Schlegel}, {Sierra-Porta}, {Tarle}, {Weaver}, {Y{\`e}che}, {Zarrouk}, {Zhou}, {Zhou}, \& {Zou}}]{2023AJ....165..253H}
{Hahn}, C., {Wilson}, M.~J., {Ruiz-Macias}, O., {et~al.} 2023, \aj, 165, 253

\bibitem[{{Heydenreich} {et~al.}(2025){Heydenreich}, {Leauthaud}, {Blake}, {Sun}, {Lange}, {Zhang}, {DeMartino}, {Ross}, {Aguilar}, {Ahlen}, {Bianchi}, {Brooks}, {Castander}, {Claybaugh}, {Cuceu}, {de la Macorra}, {DeRose}, {Dey}, {Dey}, {Doel}, {Emas}, {Ferraro}, {Font-Ribera}, {Forero-Romero}, {Garcia-Quintero}, {Gazta{\~n}aga}, {Gontcho}, {Gutierrez}, {Hadzhiyska}, {Honscheid}, {Huterer}, {Ishak}, {Jeffrey}, {Joudaki}, {Jullo}, {Juneau}, {Kirkby}, {Kisner}, {Kremin}, {Krolewski}, {Lahav}, {Lamman}, {Landriau}, {Le Guillou}, {Manera}, {Meisner}, {Miquel}, {Nadathur}, {Palanque-Delabrouille}, {Percival}, {Porredon}, {Prada}, {P{\'e}rez-R{\`a}fols}, {Rossi}, {Ruggeri}, {Sanchez}, {Saulder}, {Schlegel}, {Semenaite}, {Silber}, {Sprayberry}, {Tarl{\'e}}, {Weaver}, {Yuan}, {Zarrouk}, {Zhou}, \& {Zou}}]{Heydenreich:2025}
{Heydenreich}, S., {Leauthaud}, A., {Blake}, C., {et~al.} 2025, arXiv e-prints, arXiv:2506.21677

\bibitem[{{Heymans} {et~al.}(2021){Heymans}, {Tr{\"o}ster}, {Asgari}, {Blake}, {Hildebrandt}, {Joachimi}, {Kuijken}, {Lin}, {S{\'a}nchez}, {van den Busch}, {Wright}, {Amon}, {Bilicki}, {de Jong}, {Crocce}, {Dvornik}, {Erben}, {Fortuna}, {Getman}, {Giblin}, {Glazebrook}, {Hoekstra}, {Joudaki}, {Kannawadi}, {K{\"o}hlinger}, {Lidman}, {Miller}, {Napolitano}, {Parkinson}, {Schneider}, {Shan}, {Valentijn}, {Verdoes Kleijn}, \& {Wolf}}]{arxiv:2007.15632}
{Heymans}, C., {Tr{\"o}ster}, T., {Asgari}, M., {et~al.} 2021, \aap, 646, A140

\bibitem[{{Hildebrandt} {et~al.}(2021){Hildebrandt}, {van den Busch}, {Wright}, {Blake}, {Joachimi}, {Kuijken}, {Tr{\"o}ster}, {Asgari}, {Bilicki}, {de Jong}, {Dvornik}, {Erben}, {Getman}, {Giblin}, {Heymans}, {Kannawadi}, {Lin}, \& {Shan}}]{2021A&A...647A.124H}
{Hildebrandt}, H., {van den Busch}, J.~L., {Wright}, A.~H., {et~al.} 2021, \aap, 647, A124

\bibitem[{{Hogg}(1999)}]{1999astro.ph..5116H}
{Hogg}, D.~W. 1999, arXiv e-prints, astro

\bibitem[{{Joudaki} {et~al.}(2017){Joudaki}, {Mead}, {Blake}, {Choi}, {de Jong}, {Erben}, {Fenech Conti}, {Herbonnet}, {Heymans}, {Hildebrandt}, {Hoekstra}, {Joachimi}, {Klaes}, {K{\"o}hlinger}, {Kuijken}, {McFarland}, {Miller}, {Schneider}, \& {Viola}}]{2017MNRAS.471.1259J}
{Joudaki}, S., {Mead}, A., {Blake}, C., {et~al.} 2017, \mnras, 471, 1259

\bibitem[{{Krause} {et~al.}(2017){Krause}, {Eifler}, {Zuntz}, {Friedrich}, {Troxel}, {Dodelson}, {Blazek}, {Secco}, {MacCrann}, {Baxter}, {Chang}, {Chen}, {Crocce}, {DeRose}, {Ferte}, {Kokron}, {Lacasa}, {Miranda}, {Omori}, {Porredon}, {Rosenfeld}, {Samuroff}, {Wang}, {Wechsler}, {Abbott}, {Abdalla}, {Allam}, {Annis}, {Bechtol}, {Benoit-Levy}, {Bernstein}, {Brooks}, {Burke}, {Capozzi}, {Carrasco Kind}, {Carretero}, {D'Andrea}, {da Costa}, {Davis}, {DePoy}, {Desai}, {Diehl}, {Dietrich}, {Evrard}, {Flaugher}, {Fosalba}, {Frieman}, {Garcia-Bellido}, {Gaztanaga}, {Giannantonio}, {Gruen}, {Gruendl}, {Gschwend}, {Gutierrez}, {Honscheid}, {James}, {Jeltema}, {Kuehn}, {Kuhlmann}, {Lahav}, {Lima}, {Maia}, {March}, {Marshall}, {Martini}, {Menanteau}, {Miquel}, {Nichol}, {Plazas}, {Romer}, {Rykoff}, {Sanchez}, {Scarpine}, {Schindler}, {Schubnell}, {Sevilla-Noarbe}, {Smith}, {Soares-Santos}, {Sobreira}, {Suchyta}, {Swanson}, {Tarle}, {Tucker}, {Vikram}, {Walker}, \& {Weller}}]{Krause:2017}
{Krause}, E., {Eifler}, T.~F., {Zuntz}, J., {et~al.} 2017, arXiv e-prints, arXiv:1706.09359

\bibitem[{{Lange}(2024)}]{Lange:2024}
{Lange}, J.~U. 2024, in American Astronomical Society Meeting Abstracts, Vol. 244, American Astronomical Society Meeting Abstracts, 229.02

\bibitem[{{Lange} {et~al.}(2022){Lange}, {Hearin}, {Leauthaud}, {van den Bosch}, {Guo}, \& {DeRose}}]{2022MNRAS.509.1779L}
{Lange}, J.~U., {Hearin}, A.~P., {Leauthaud}, A., {et~al.} 2022, \mnras, 509, 1779

\bibitem[{{Lange} {et~al.}(2023){Lange}, {Hearin}, {Leauthaud}, {van den Bosch}, {Xhakaj}, {Guo}, {Wechsler}, \& {DeRose}}]{2023MNRAS.520.5373L}
{Lange}, J.~U., {Hearin}, A.~P., {Leauthaud}, A., {et~al.} 2023, \mnras, 520, 5373

\bibitem[{{Lange} {et~al.}(2021){Lange}, {Leauthaud}, {Singh}, {Guo}, {Zhou}, {Smith}, \& {Cyr-Racine}}]{2021MNRAS.502.2074L}
{Lange}, J.~U., {Leauthaud}, A., {Singh}, S., {et~al.} 2021, \mnras, 502, 2074

\bibitem[{{Lange} {et~al.}(2019{\natexlab{a}}){Lange}, {van den Bosch}, {Zentner}, {Wang}, {Hearin}, \& {Guo}}]{Lange:2019b}
{Lange}, J.~U., {van den Bosch}, F.~C., {Zentner}, A.~R., {et~al.} 2019{\natexlab{a}}, \mnras, 490, 1870

\bibitem[{{Lange} {et~al.}(2019{\natexlab{b}}){Lange}, {Yang}, {Guo}, {Luo}, \& {van den Bosch}}]{Lange:2019a}
{Lange}, J.~U., {Yang}, X., {Guo}, H., {Luo}, W., \& {van den Bosch}, F.~C. 2019{\natexlab{b}}, \mnras, 488, 5771

\bibitem[{{Leauthaud} {et~al.}(2022){Leauthaud}, {Amon}, {Singh}, {Gruen}, {Lange}, {Huang}, {Robertson}, {Varga}, {Luo}, {Heymans}, {Hildebrandt}, {Blake}, {Aguena}, {Allam}, {Andrade-Oliveira}, {Annis}, {Bertin}, {Bhargava}, {Blazek}, {Bridle}, {Brooks}, {Burke}, {Rosell}, {Carrasco Kind}, {Carretero}, {Castander}, {Cawthon}, {Choi}, {Costanzi}, {da Costa}, {Pereira}, {Davis}, {De Vicente}, {DeRose}, {Diehl}, {Dietrich}, {Doel}, {Eckert}, {Everett}, {Evrard}, {Ferrero}, {Flaugher}, {Fosalba}, {Garc{\'\i}a-Bellido}, {Gatti}, {Gaztanaga}, {Gruendl}, {Gschwend}, {Hartley}, {Hollowood}, {Honscheid}, {Jain}, {James}, {Jarvis}, {Joachimi}, {Kannawadi}, {Kim}, {Krause}, {Kuehn}, {Kuijken}, {Kuropatkin}, {Lima}, {MacCrann}, {Maia}, {Makler}, {March}, {Marshall}, {Melchior}, {Menanteau}, {Miquel}, {Miyatake}, {Mohr}, {Moraes}, {More}, {Surhud}, {Morgan}, {Myles}, {Ogando}, {Palmese}, {Paz-Chinch{\'o}n}, {Malag{\'o}n}, {Prat}, {Rau}, {Rhodes}, {Rodriguez-Monroy}, {Roodman}, {Ross}, {Samuroff}, {S{\'a}nchez}, {Sanchez}, {Scarpine}, {Schlegel}, {Schubnell}, {Serrano}, {Sevilla-Noarbe}, {Sif{\'o}n}, {Smith}, {Speagle}, {Suchyta}, {Tarle}, {Thomas}, {Tinker}, {To}, {Troxel}, {Van Waerbeke}, {Vielzeuf}, \& {Wright}}]{Leauthaud:2022}
{Leauthaud}, A., {Amon}, A., {Singh}, S., {et~al.} 2022, \mnras, 510, 6150

\bibitem[{{Leauthaud} {et~al.}(2017){Leauthaud}, {Saito}, {Hilbert}, {Barreira}, {More}, {White}, {Alam}, {Behroozi}, {Bundy}, {Coupon}, {Erben}, {Heymans}, {Hildebrandt}, {Mandelbaum}, {Miller}, {Moraes}, {Pereira}, {Rodr{\'\i}guez-Torres}, {Schmidt}, {Shan}, {Viel}, \& {Villaescusa-Navarro}}]{Leauthaud:2017}
{Leauthaud}, A., {Saito}, S., {Hilbert}, S., {et~al.} 2017, \mnras, 467, 3024

\bibitem[{{Li} {et~al.}(2023){Li}, {Zhang}, {Sugiyama}, {Dalal}, {Terasawa}, {Rau}, {Mandelbaum}, {Takada}, {More}, {Strauss}, {Miyatake}, {Shirasaki}, {Hamana}, {Oguri}, {Luo}, {Nishizawa}, {Takahashi}, {Nicola}, {Osato}, {Kannawadi}, {Sunayama}, {Armstrong}, {Bosch}, {Komiyama}, {Lupton}, {Lust}, {MacArthur}, {Miyazaki}, {Murayama}, {Nishimichi}, {Okura}, {Price}, {Tait}, {Tanaka}, \& {Wang}}]{2023PhRvD.108l3518L}
{Li}, X., {Zhang}, T., {Sugiyama}, S., {et~al.} 2023, \prd, 108, 123518

\bibitem[{{Limber}(1953)}]{Limber:1953}
{Limber}, D.~N. 1953, \apj, 117, 134

\bibitem[{{Linder}(2003)}]{2003PhRvL..90i1301L}
{Linder}, E.~V. 2003, \prl, 90, 091301

\bibitem[{{Lodha} {et~al.}(2025){Lodha}, {Calderon}, {Matthewson}, {Shafieloo}, {Ishak}, {Pan}, {Garcia-Quintero}, {Huterer}, {Valogiannis}, {Ure{\~n}a-L{\'o}pez}, {Kamble}, {Parkinson}, {Kim}, {Zhao}, {Cervantes-Cota}, {Rohlf}, {Lozano-Rodr{\'\i}guez}, {Rom{\'a}n-Herrera}, {Abdul-Karim}, {Aguilar}, {Ahlen}, {Alves}, {Andrade}, {Armengaud}, {Aviles}, {BenZvi}, {Bianchi}, {Brodzeller}, {Brooks}, {Burtin}, {Canning}, {Carnero Rosell}, {Casas}, {Castander}, {Charles}, {Chaussidon}, {Chaves-Montero}, {Chebat}, {Claybaugh}, {Cole}, {Cuceu}, {Dawson}, {de la Macorra}, {de Mattia}, {Deiosso}, {Demina}, {Dey}, {Dey}, {Ding}, {Doel}, {Eisenstein}, {Elbers}, {Ferraro}, {Font-Ribera}, {Forero-Romero}, {Garrison}, {Gazta{\~n}aga}, {Gil-Mar{\'\i}n}, {Gontcho}, {Gonzalez-Morales}, {Gutierrez}, {Guy}, {Hahn}, {Herbold}, {Herrera-Alcantar}, {Honscheid}, {Howlett}, {Juneau}, {Kehoe}, {Kirkby}, {Kisner}, {Kremin}, {Lahav}, {Lamman}, {Landriau}, {Le Guillou}, {Leauthaud}, {Levi}, {Li}, {Magneville}, {Manera}, {Martini}, {Meisner}, {Mena-Fern{\'a}ndez}, {Miquel}, {Moustakas}, {Mu{\~n}oz Santos}, {Mu{\~n}oz-Guti{\'e}rrez}, {Myers}, {Nadathur}, {Niz}, {Noriega}, {Paillas}, {Palanque-Delabrouille}, {Percival}, {Pieri}, {Poppett}, {Prada}, {P{\'e}rez-Fern{\'a}ndez}, {P{\'e}rez-R{\`a}fols}, {Ram{\'\i}rez-P{\'e}rez}, {Rashkovetskyi}, {Ravoux}, {Ross}, {Rossi}, {Ruhlmann-Kleider}, {Samushia}, {Sanchez}, {Schlegel}, {Schubnell}, {Seo}, {Sinigaglia}, {Sprayberry}, {Tan}, {Tarl{\'e}}, {Taylor}, {Turner}, {Vargas-Maga{\~n}a}, {Walther}, {Weaver}, {Wolfson}, {Y{\`e}che}, {Zarrouk}, {Zhou}, \& {Zou}}]{arXiv:2503.14743}
{Lodha}, K., {Calderon}, R., {Matthewson}, W.~L., {et~al.} 2025, arXiv e-prints, arXiv:2503.14743

\bibitem[{{Maksimova} {et~al.}(2021){Maksimova}, {Garrison}, {Eisenstein}, {Hadzhiyska}, {Bose}, \& {Satterthwaite}}]{2021MNRAS.508.4017M}
{Maksimova}, N.~A., {Garrison}, L.~H., {Eisenstein}, D.~J., {et~al.} 2021, \mnras, 508, 4017

\bibitem[{{Miyatake} {et~al.}(2023){Miyatake}, {Sugiyama}, {Takada}, {Nishimichi}, {Li}, {Shirasaki}, {More}, {Kobayashi}, {Nishizawa}, {Rau}, {Zhang}, {Takahashi}, {Dalal}, {Mandelbaum}, {Strauss}, {Hamana}, {Oguri}, {Osato}, {Luo}, {Kannawadi}, {Hsieh}, {Armstrong}, {Bosch}, {Komiyama}, {Lupton}, {Lust}, {MacArthur}, {Miyazaki}, {Murayama}, {Okura}, {Price}, {Sunayama}, {Tait}, {Tanaka}, \& {Wang}}]{arxiv:2304.00704}
{Miyatake}, H., {Sugiyama}, S., {Takada}, M., {et~al.} 2023, \prd, 108, 123517

\bibitem[{{Myles} {et~al.}(2021){Myles}, {Alarcon}, {Amon}, {S{\'a}nchez}, {Everett}, {DeRose}, {McCullough}, {Gruen}, {Bernstein}, {Troxel}, {Dodelson}, {Campos}, {MacCrann}, {Yin}, {Raveri}, {Amara}, {Becker}, {Choi}, {Cordero}, {Eckert}, {Gatti}, {Giannini}, {Gschwend}, {Gruendl}, {Harrison}, {Hartley}, {Huff}, {Kuropatkin}, {Lin}, {Masters}, {Miquel}, {Prat}, {Roodman}, {Rykoff}, {Sevilla-Noarbe}, {Sheldon}, {Wechsler}, {Yanny}, {Abbott}, {Aguena}, {Allam}, {Annis}, {Bacon}, {Bertin}, {Bhargava}, {Bridle}, {Brooks}, {Burke}, {Carnero Rosell}, {Carrasco Kind}, {Carretero}, {Castander}, {Conselice}, {Costanzi}, {Crocce}, {da Costa}, {Pereira}, {Desai}, {Diehl}, {Eifler}, {Elvin-Poole}, {Evrard}, {Ferrero}, {Fert{\'e}}, {Flaugher}, {Fosalba}, {Frieman}, {Garc{\'\i}a-Bellido}, {Gaztanaga}, {Giannantonio}, {Hinton}, {Hollowood}, {Honscheid}, {Hoyle}, {Huterer}, {James}, {Krause}, {Kuehn}, {Lahav}, {Lima}, {Maia}, {Marshall}, {Martini}, {Melchior}, {Menanteau}, {Mohr}, {Morgan}, {Muir}, {Ogando}, {Palmese}, {Paz-Chinch{\'o}n}, {Plazas}, {Rodriguez-Monroy}, {Samuroff}, {Sanchez}, {Scarpine}, {Secco}, {Serrano}, {Smith}, {Soares-Santos}, {Suchyta}, {Swanson}, {Tarle}, {Thomas}, {To}, {Varga}, {Weller}, \& {Wester}}]{2021MNRAS.505.4249M}
{Myles}, J., {Alarcon}, A., {Amon}, A., {et~al.} 2021, \mnras, 505, 4249

\bibitem[{{Neistein} {et~al.}(2011){Neistein}, {Li}, {Khochfar}, {Weinmann}, {Shankar}, \& {Boylan-Kolchin}}]{Neistein:2011}
{Neistein}, E., {Li}, C., {Khochfar}, S., {et~al.} 2011, \mnras, 416, 1486

\bibitem[{{Planck Collaboration} {et~al.}(2020){Planck Collaboration}, {Aghanim}, {Akrami}, {Ashdown}, {Aumont}, {Baccigalupi}, {Ballardini}, {Banday}, {Barreiro}, {Bartolo}, {Basak}, {Battye}, {Benabed}, {Bernard}, {Bersanelli}, {Bielewicz}, {Bock}, {Bond}, {Borrill}, {Bouchet}, {Boulanger}, {Bucher}, {Burigana}, {Butler}, {Calabrese}, {Cardoso}, {Carron}, {Challinor}, {Chiang}, {Chluba}, {Colombo}, {Combet}, {Contreras}, {Crill}, {Cuttaia}, {de Bernardis}, {de Zotti}, {Delabrouille}, {Delouis}, {Di Valentino}, {Diego}, {Dor{\'e}}, {Douspis}, {Ducout}, {Dupac}, {Dusini}, {Efstathiou}, {Elsner}, {En{\ss}lin}, {Eriksen}, {Fantaye}, {Farhang}, {Fergusson}, {Fernandez-Cobos}, {Finelli}, {Forastieri}, {Frailis}, {Fraisse}, {Franceschi}, {Frolov}, {Galeotta}, {Galli}, {Ganga}, {G{\'e}nova-Santos}, {Gerbino}, {Ghosh}, {Gonz{\'a}lez-Nuevo}, {G{\'o}rski}, {Gratton}, {Gruppuso}, {Gudmundsson}, {Hamann}, {Handley}, {Hansen}, {Herranz}, {Hildebrandt}, {Hivon}, {Huang}, {Jaffe}, {Jones}, {Karakci}, {Keih{\"a}nen}, {Keskitalo}, {Kiiveri}, {Kim}, {Kisner}, {Knox}, {Krachmalnicoff}, {Kunz}, {Kurki-Suonio}, {Lagache}, {Lamarre}, {Lasenby}, {Lattanzi}, {Lawrence}, {Le Jeune}, {Lemos}, {Lesgourgues}, {Levrier}, {Lewis}, {Liguori}, {Lilje}, {Lilley}, {Lindholm}, {L{\'o}pez-Caniego}, {Lubin}, {Ma}, {Mac{\'\i}as-P{\'e}rez}, {Maggio}, {Maino}, {Mandolesi}, {Mangilli}, {Marcos-Caballero}, {Maris}, {Martin}, {Martinelli}, {Mart{\'\i}nez-Gonz{\'a}lez}, {Matarrese}, {Mauri}, {McEwen}, {Meinhold}, {Melchiorri}, {Mennella}, {Migliaccio}, {Millea}, {Mitra}, {Miville-Desch{\^e}nes}, {Molinari}, {Montier}, {Morgante}, {Moss}, {Natoli}, {N{\o}rgaard-Nielsen}, {Pagano}, {Paoletti}, {Partridge}, {Patanchon}, {Peiris}, {Perrotta}, {Pettorino}, {Piacentini}, {Polastri}, {Polenta}, {Puget}, {Rachen}, {Reinecke}, {Remazeilles}, {Renzi}, {Rocha}, {Rosset}, {Roudier}, {Rubi{\~n}o-Mart{\'\i}n}, {Ruiz-Granados}, {Salvati}, {Sandri}, {Savelainen}, {Scott}, {Shellard}, {Sirignano}, {Sirri}, {Spencer}, {Sunyaev}, {Suur-Uski}, {Tauber}, {Tavagnacco}, {Tenti}, {Toffolatti}, {Tomasi}, {Trombetti}, {Valenziano}, {Valiviita}, {Van Tent}, {Vibert}, {Vielva}, {Villa}, {Vittorio}, {Wandelt}, {Wehus}, {White}, {White}, {Zacchei}, \& {Zonca}}]{2020A&A...641A...6P}
{Planck Collaboration}, {Aghanim}, N., {Akrami}, Y., {et~al.} 2020, \aap, 641, A6

\bibitem[{{Rau} {et~al.}(2023){Rau}, {Dalal}, {Zhang}, {Li}, {Nishizawa}, {More}, {Mandelbaum}, {Miyatake}, {Strauss}, \& {Takada}}]{2023MNRAS.524.5109R}
{Rau}, M.~M., {Dalal}, R., {Zhang}, T., {et~al.} 2023, \mnras, 524, 5109

\bibitem[{{Sailer} {et~al.}(2025){Sailer}, {DeRose}, {Ferraro}, {Chen}, {Zhou}, {White}, {Kim}, \& {Madhavacheril}}]{2025PhRvD.111j3540S}
{Sailer}, N., {DeRose}, J., {Ferraro}, S., {et~al.} 2025, \prd, 111, 103540

\bibitem[{{Scherrer}(2015)}]{arXiv:1505.05781}
{Scherrer}, R.~J. 2015, \prd, 92, 043001

\bibitem[{{Stebbins}(1996)}]{Stebbins:1996}
{Stebbins}, A. 1996, arXiv e-prints, astro

\bibitem[{{Sunyaev} \& {Zeldovich}(1972)}]{1972CoASP...4..173S}
{Sunyaev}, R.~A. \& {Zeldovich}, Y.~B. 1972, Comments on Astrophysics and Space Physics, 4, 173

\bibitem[{{Terasawa} {et~al.}(2025){Terasawa}, {Li}, {Takada}, {Nishimichi}, {Tanaka}, {Sugiyama}, {Kurita}, {Zhang}, {Shirasaki}, {Takahashi}, {Miyatake}, {More}, \& {Nishizawa}}]{arxiv:2403.20323}
{Terasawa}, R., {Li}, X., {Takada}, M., {et~al.} 2025, \prd, 111, 063509

\bibitem[{{Wright} {et~al.}(2025){Wright}, {St{\"o}lzner}, {Asgari}, {Bilicki}, {Giblin}, {Heymans}, {Hildebrandt}, {Hoekstra}, {Joachimi}, {Kuijken}, {Li}, {Reischke}, {von Wietersheim-Kramsta}, {Yoon}, {Burger}, {Chisari}, {de Jong}, {Dvornik}, {Georgiou}, {Harnois-D{\'e}raps}, {Jalan}, {William}, {Joudaki}, {Lesci}, {Linke}, {Loureiro}, {Mahony}, {Maturi}, {Miller}, {Moscardini}, {Napolitano}, {Porth}, {Radovich}, {Schneider}, {Tr{\"o}ster}, {Wittje}, {Yan}, \& {Zhang}}]{2025arXiv250319441W}
{Wright}, A.~H., {St{\"o}lzner}, B., {Asgari}, M., {et~al.} 2025, arXiv e-prints, arXiv:2503.19441

\bibitem[{{Yuan} {et~al.}(2024){Yuan}, {Zhang}, {Ross}, {Donald-McCann}, {Hadzhiyska}, {Wechsler}, {Zheng}, {Alam}, {Gonzalez-Perez}, {Aguilar}, {Ahlen}, {Bianchi}, {Brooks}, {de la Macorra}, {Fanning}, {Forero-Romero}, {Honscheid}, {Ishak}, {Kehoe}, {Lasker}, {Landriau}, {Manera}, {Martini}, {Meisner}, {Miquel}, {Moustakas}, {Nadathur}, {Newman}, {Nie}, {Percival}, {Poppett}, {Rocher}, {Rossi}, {Sanchez}, {Samushia}, {Schubnell}, {Seo}, {Tarl{\'e}}, {Weaver}, {Yu}, {Zhou}, \& {Zou}}]{Yuan:2024}
{Yuan}, S., {Zhang}, H., {Ross}, A.~J., {et~al.} 2024, \mnras, 530, 947

\bibitem[{{Zheng} {et~al.}(2005){Zheng}, {Berlind}, {Weinberg}, {Benson}, {Baugh}, {Cole}, {Dav{\'e}}, {Frenk}, {Katz}, \& {Lacey}}]{2005ApJ...633..791Z}
{Zheng}, Z., {Berlind}, A.~A., {Weinberg}, D.~H., {et~al.} 2005, \apj, 633, 791

\bibitem[{{Zheng} \& {Guo}(2016)}]{Zheng:2016}
{Zheng}, Z. \& {Guo}, H. 2016, \mnras, 458, 4015

\bibitem[{{Zhou} {et~al.}(2023){Zhou}, {Dey}, {Newman}, {Eisenstein}, {Dawson}, {Bailey}, {Berti}, {Guy}, {Lan}, {Zou}, {Aguilar}, {Ahlen}, {Alam}, {Brooks}, {de la Macorra}, {Dey}, {Dhungana}, {Fanning}, {Font-Ribera}, {Gontcho}, {Honscheid}, {Ishak}, {Kisner}, {Kov{\'a}cs}, {Kremin}, {Landriau}, {Levi}, {Magneville}, {Manera}, {Martini}, {Meisner}, {Miquel}, {Moustakas}, {Myers}, {Nie}, {Palanque-Delabrouille}, {Percival}, {Poppett}, {Prada}, {Raichoor}, {Ross}, {Schlafly}, {Schlegel}, {Schubnell}, {Tarl{\'e}}, {Weaver}, {Wechsler}, {Y{\'e}che}, \& {Zhou}}]{2023AJ....165...58Z}
{Zhou}, R., {Dey}, B., {Newman}, J.~A., {et~al.} 2023, \aj, 165, 58

\bibitem[{{Zu}(2020)}]{arxiv:2010.01143}
{Zu}, Y. 2020, arXiv e-prints, arXiv:2010.01143

\end{thebibliography}
\end{document}